%% file: DvsMultpPb.tex
\documentclass[ALICE,manyauthors]{cernphprep}
\pdfoutput=1
\usepackage{hyperref}
\usepackage{lineno}

\usepackage[english]{babel}
\usepackage{graphicx}
\usepackage{verbatim}
\usepackage{amsfonts}
\usepackage{makeidx}
\usepackage{color}
\usepackage{amsmath}
\usepackage{epstopdf}
\usepackage{hyperref}
\usepackage{cite}
\usepackage{placeins}
\usepackage{subfigure}
\usepackage{rotating}
\usepackage{lscape}
\usepackage{multirow}

\usepackage[comma,square,numbers,sort&compress]{natbib}

\newcommand{\pp}{\text{pp}}
\newcommand{\ppbar}{\text{p}\overline{\text{p}}}

\newcommand{\pPb}{\mbox{p--Pb}}
\newcommand{\sqrts}{\sqrt{s}}
\newcommand{\sqrtsNN}{\sqrt{s_{\rm NN}}}

\newcommand{\av}[1]{\left\langle #1 \right\rangle}

\newcommand{\gev}{\mathrm{GeV}}
\newcommand{\gevc}{\ensuremath{\mathrm{GeV}/c}}
\newcommand{\tev}{\mathrm{TeV}}

\newcommand{\mb}{\mathrm{mb}}

\newcommand{\pt}{p_{\rm T}}

\newcommand{\ylab}{y_{\rm lab}}

\newcommand{\DtoKpi}{{\rm D}^0 \to {\rm K}^-\pi^+}
\newcommand{\DtoKpipi}{{\rm D}^+\to {\rm K}^-\pi^+\pi^+}
\newcommand{\DstartoDpi}{{\rm D}^{*+} \to {\rm D}^0 \pi^+}

\newcommand{\Dzero}{{\rm D^0}}

\newcommand{\Dstar}{{\rm D^{*+}}}

\newcommand{\Dplus}{{\rm D^+}}

\newcommand{\Jpsi}{{\rm J}/\psi}

\newcommand{\Ntrk}{N_{\rm tracklets}}
\newcommand{\Nch}{N_{\rm ch}}
\newcommand{\zvtx}{z_{\rm vtx}}

\newcommand{\Nvzero}{N_{\rm V0A}}
\newcommand{\dNdEta}{{\rm d}N_{\rm ch}/{\rm d}\eta}
\newcommand{\dNdpt}{{\rm d}N^{\rm D}/{\rm d}\pt}

\newcommand{\Ncoll}{\ensuremath{N_{\rm coll}}}
\newcommand{\Npart}{\ensuremath{N_{\rm part}}}
\newcommand{\Ncollmult}{\ensuremath{N_{\rm coll}^\mathrm{mult}}}
\newcommand{\Npartmult}{\ensuremath{N_{\rm part}^\mathrm{mult}}}
\newcommand{\Ncollglauber}{\ensuremath{N_{\rm coll}^\mathrm{Glauber}}}
\newcommand{\Npartglauber}{\ensuremath{N_{\rm part}^\mathrm{Glauber}}}

\newcommand{\fprompt}{\ensuremath{f_\mathrm{prompt}}}
\newcommand{\RpPb}{\ensuremath{R_\mathrm{pPb}}}
\newcommand{\QpPb}{\ensuremath{Q_\mathrm{pPb}}}
\newcommand{\TpPb}{\ensuremath{T_\mathrm{pPb}}}

\newcommand{\TpPbmult}{\ensuremath{T_\mathrm{pPb}^\mathrm{mult}}}

\newcommand{\TpPbglauber}{\ensuremath{T_\mathrm{pPb}^\mathrm{Glauber}}}

\newcommand{\CL}{{CL1}}
\newcommand{\VzA}{{V0A}}
\newcommand{\ZNA}{{ZNA}}

\begin{document}%

\begin{titlepage}
\PHyear{2016}
\PHnumber{034}      
\PHdate{15 February}  

\title{Measurement of D-meson production versus multiplicity in \pPb~collisions at ${\mathbf \sqrtsNN=5.02}$~TeV}
\ShortTitle{D-meson production versus multiplicity in \pPb~collisions}   

\Collaboration{ALICE Collaboration\thanks{See Appendix~\ref{app:collab} for the list of collaboration members}}
\ShortAuthor{ALICE Collaboration} 

\begin{abstract}

The measurement of prompt D-meson production as a function of multiplicity in \pPb~collisions at $\sqrtsNN=5.02$~TeV with the ALICE detector at the LHC is reported.  
$\Dzero$, $\Dplus$ and $\Dstar$ mesons are reconstructed via their hadronic decay channels in the centre-of-mass rapidity range $-0.96< y_{\mathrm{cms}}<0.04$ and transverse momentum interval $1<\pt<24~\gevc$. 
The multiplicity dependence of D-meson production is examined by either comparing yields in \pPb~collisions in different event classes, selected based on the multiplicity of produced particles or zero-degree energy, with those in \pp~collisions, scaled by the number of binary nucleon--nucleon collisions (nuclear modification factor);  
as well as by evaluating the per-event yields in \pPb~collisions in different multiplicity intervals normalised to the multiplicity-integrated ones (relative yields).  
The nuclear modification factors for $\Dzero$, $\Dplus$ and $\Dstar$ are consistent with one another. The D-meson nuclear modification factors as a function of the zero-degree energy are consistent with unity within uncertainties in the measured $\pt$ regions and event classes.  
The relative D-meson yields, calculated in various $\pt$ intervals, increase as a function of the charged-particle multiplicity. The results are compared with the equivalent \pp~measurements at $\sqrt{s}=7$~TeV as well as with EPOS~3 calculations. 
\end{abstract}

\end{titlepage}
\setcounter{page}{2}

\section{Introduction}
\label{sec:intro}

%

In high-energy hadronic collisions, heavy quarks (charm and beauty) are produced in hard parton scattering processes.  
Due to their large masses, their production cross sections can be calculated in the framework of perturbative Quantum Chromodynamics (pQCD) down to low transverse momenta.  
The differential cross section for heavy-flavour hadron production in nucleon-nucleon collisions can be calculated in the factorisation approach by the convolution of parton densities in the incoming nucleon, the short-distance partonic cross section of heavy quark production, and the fragmentation function that describes the transition of the heavy quark into a heavy-flavour hadron~\cite{Andronic:2015wma}.  
Thus, heavy-flavour production is sensitive to the gluon and the possible heavy-quark content in the nucleon and provides constraints on the parton distribution functions (PDFs) in the proton and in the nucleus~\cite{Kniehl:2009ar,Stavreva:2010mw}. Measurements of heavy-flavour hadron production in hadronic collisions provide tests of pQCD and constitute a crucial baseline for the study of heavy-flavour production in heavy-ion collisions~\cite{ALICE:2012ab,Abelev:2012qh}.  
A suppression of heavy-flavour yields is observed in heavy-ion collisions at high transverse momentum ($\pt$), and is interpreted as being due to the formation of a Quark-Gluon Plasma (QGP). 
 
Beauty production measurements in $\ppbar$~collisions at $\sqrts = 1.96$~TeV at the FNAL Tevatron collider~\cite{Acosta:2004yw,Cacciari:2003uh,Kniehl:2008zza} and in $\pp$~collisions at $\sqrts = 7$~TeV at the CERN LHC collider~\cite{Abelev:2012gx,Abelev:2012sca,ATLAS:2013cia,Khachatryan:2010yr,Aaij:2010gn} are described by different implementations of pQCD calculations, such as the General-Mass-Variable-Flavour-Number Scheme (GM-VFNS)~\cite{Kniehl:2004fy,Kniehl:2005mk} at next-to-leading order, and the Fixed-Order plus Next-to-Leading Logarithms (FONLL) approach~\cite{Cacciari:1998it,Cacciari:2001td,Cacciari:2012ny}.  
Analogously, inclusive charm meson production measurements at the LHC~\cite{ALICE:2011aa,Abelev:2012vra,Aaij:2013mga} are reproduced within uncertainties by the predictions of GM-VFNS, FONLL and those performed in the framework of $k_{\rm T}$ factorisation in the Leading Order (LO) approximation~\cite{Maciula:2013wg}.

Recently, the study of heavy-flavour production as a function of the multiplicity of charged particles produced in the collision has attracted growing interest.  
Such measurements probe the interplay between hard and soft mechanisms in particle production.  
At LHC energies, the multiplicity dependence of heavy-flavour production is likely to be affected by the larger amount of gluon radiation associated with short-distance production processes, as well as by the contribution of Multiple-Parton Interactions (MPI)~\cite{Bartalini:2010su,Sjostrand:1987su,Porteboeuf:2010dw}.  
It has also been argued that, due to the spatial distribution of partons in the transverse plane, the probability for MPI to occur in a pp collision increases towards smaller impact parameters~\cite{Frankfurt:2010ea,Azarkin:2014cja,Strikman:2011zz}.  
This effect might be further enhanced by quantum-mechanical fluctuations of gluon densities at small Bjorken-$x$~\cite{Strikman:2011ar}. 
 
The measurements of prompt D mesons, inclusive and non-prompt $\Jpsi$ in $\pp$~collisions at $\sqrts= 7$~TeV~\cite{Abelev:2012rz,Adam:2015ota}, and of the three $\Upsilon$ states in $\pp$~collisions at $\sqrts= 2.76$~TeV~\cite{Chatrchyan:2013nza},  
provide evidence for a similar increase of open and hidden heavy-flavour yields as a function of charged-particle multiplicity. These results suggest that the enhancement probably originates in short-distance production processes, and is not influenced by hadronisation mechanisms.  
The enhancement is quantitatively described by calculations including MPI contributions, namely percolation model estimates~\cite{Ferreiro:2012fb,Ferreiro:2015gea}, the EPOS~3 event generator~\cite{Drescher:2000ha,Werner:2013tya} and PYTHIA 8.157 calculations~\cite{Sjostrand:2007gs}.

In proton-nucleus collisions, several so-called `Cold Nuclear Matter' (CNM) effects occur due to the presence of a nucleus in the colliding system, and, possibly, to the large density of produced particles. These CNM effects can affect the production of heavy-flavour hadrons at all the stages of their formation. 
In particular, the PDFs of nucleons bound in nuclei are modified with respect to those of free nucleons.  
This modification of the PDFs in the nucleus can be described by phenomenological parameterisations (nuclear PDFs, or nPDFs)~\cite{Eskola:2009uj,deFlorian:2003qf,Hirai:2007sx}. Alternatively, when the production process is dominated by 
gluons at low Bjorken-$x$, the nucleus can be described by the Colour-Glass Condensate (CGC) effective theory as a coherent and saturated gluonic system~\cite{Fujii:2013yja,Tribedy:2011aa,Albacete:2012xq,Rezaeian:2012ye}.  
The kinematics of the partons in the initial state can be affected by multiple scatterings (transverse momentum broadening, or $k_\mathrm{T}$ broadening)~\cite{Lev:1983hh,Wang:1998ww,Kopeliovich:2002yh} or by gluon radiation (energy loss)~\cite{Vitev:2007ve} before the heavy-quark pair is produced. Gluon radiation may also occur after the heavy-quark pair is formed~\cite{Arleo:2010rb}.  
Other measurements in \pPb~collisions at $\sqrtsNN = 5.02$~TeV, e.g.\ those of angular correlations between charged particles~\cite{CMS:2012qk,Abelev:2012ola,ABELEV:2013wsa,Aad:2012gla}, of $\psi$(2S) suppression~\cite{Abelev:2014zpa} and of the relative yields of the three $\Upsilon$ states~\cite{Chatrchyan:2013nza}, indicate that final-state effects also play an important role.

The measured charm production cross section in minimum-bias \pPb~collisions at $\sqrtsNN = 5.02$~TeV~\cite{Abelev:2014hha} is consistent within uncertainties with that in \pp~collisions at the same energy scaled by the atomic mass number of the Pb nucleus. The nuclear modification factor was also found to be consistent with calculations considering EPS09 nPDFs~\cite{Eskola:2009uj}, CGC, or transverse momentum broadening and initial-state energy loss. The influence of cold nuclear matter effects on multiplicity-integrated D-meson production in \pPb~collisions is smaller than the measurement uncertainties.

Additional insight into CNM effects can be obtained by  
measuring the heavy-flavour hadron yields as a function of the 
multiplicity of charged particles produced in the \pPb~collision. 
The aim of these studies is to explore the dependence of heavy-flavour  
production on the collision geometry and on the density of final-state  
particles. 
Indeed, it is expected that the multiplicity of produced particles depends on the number of nucleons overlapping in the collision region, and therefore on the geometry of the collision (i.e. on the collision centrality). 
 
Most of the aforementioned models of CNM effects consider a dependence on the collision geometry, usually expressed through the impact parameter of the collision, the number of participant nucleons ($\Npart$), or the number of nucleon-nucleon collisions ($\Ncoll$). 
In general, CNM effects are expected to be more pronounced in central collisions, i.e.\ those having a small impact parameter. 
Some of the parameterisations of the nPDFs have studied the influence of the local nucleon density~\cite{Helenius:2014jza,Helenius:2012wd,Emel'yanov:1998df,Emel'yanov:1999bn}.  
The spatially dependent EPS09 and EKS98 nPDF sets, EPS09s and EKS98s, are formulated as a function of the nuclear thickness~\cite{Helenius:2014jza}.  
The leading twist nuclear shadowing calculation~\cite{Frankfurt:2011cs} assumes the Glauber-Gribov approach of the collision geometry and predicts the dependence of the nPDF on the collision impact parameter.  
The estimates of the initial-state $k_{\rm T}$ broadening due to multiple soft collisions also consider a dependence on the collision impact parameter~\cite{Kopeliovich:2002yh,Wang:1998ww}.  
Initial-state parton energy loss is also expected to evolve with the collision geometry as a consequence of the different nuclear density, though detailed calculations including this effect are not yet available.  
Finally, if final-state effects were to affect heavy-flavour production in \pPb~collisions, their influence would also vary with the density of produced particles.  
 
In this paper, we report the $\pt$-differential measurements of $\Dzero$, $\Dplus$ and $\Dstar$ production as a function of multiplicity in \pPb~collisions at $\sqrtsNN = 5.02$~TeV. 
The experimental setup and the data sample are described in Sec.~\ref{sec:detector}.  
The determination of the multiplicity and the estimation of the collision centrality and of the number of nucleon-nucleon collisions are discussed in Sec.~\ref{sec:AnalysisMethods}.  
The D-meson reconstruction strategy is explained in Sec.~\ref{sec:Dreco}.  
The results are reported in the form of  
the D-meson nuclear modification factor in different centrality classes (Sec.~\ref{sec:EA}),  
and the relative D-meson yields as a function of the relative charged-particle multiplicity at central and backward rapidity (Sec.~\ref{sec:DvM}).

\section{Experimental apparatus and data sample}
\label{sec:detector}

The ALICE apparatus is described in detail in~\cite{Aamodt:2008zz} and its performance in~\cite{Abelev:2014ffa}. It is composed of a series of detectors in the central barrel for tracking and particle identification;  
the Muon Spectrometer in the forward direction for muon tracking and identification; and a further set of detectors at forward rapidity for triggering and event characterisation.  
The central barrel detectors are located inside a large solenoid magnet that provides a 0.5 T field parallel to the beam direction, which corresponds to the $z$-axis of the ALICE coordinate system.  
In this Section, the detectors used for the D-meson analysis are briefly described. 
 
The Inner Tracking System (ITS), the Time Projection Chamber (TPC) and the Time Of Flight detector (TOF) allow the reconstruction and identification of charged particles in the central pseudorapidity region. The V0 detector, composed of two scintillator arrays located in the forward and backward pseudorapidity regions, is used for online event triggering and multiplicity determination.  
The Zero Degree Calorimeters (ZDC) are used for event selection and 
to estimate the collision centrality via the zero-degree energy. 
 
The ITS is composed of six cylindrical layers of silicon detectors, located at radii between 3.9 cm (about 1 cm from the beam vacuum tube) and 43.0 cm. The two innermost layers, which respectively cover $|\eta|<2.0$ and $|\eta|<1.4$, comprise the Silicon Pixel Detectors (SPD); the two intermediate layers, within $|\eta|<0.9$, consist of Silicon Drift Detectors (SDD); and the two outer layers, also covering $|\eta|<0.9$, consist of double-sided Silicon Strip Detectors (SSD).  
The low material budget, high spatial resolution, and position of the detector setup surrounding the beam vacuum tube and close to the interaction point allow it to provide a measurement of the charged-particle impact parameter in the transverse plane ($d_0$), i.e.\ the distance of closest approach between the track and the primary vertex along $r\phi$, with a resolution better than $75~\mu$m for transverse momenta $\pt>1~\gevc$~\cite{Aamodt:2010aa}.

The TPC is a large cylindrical drift detector, extending from 85 cm to 247 cm in the radial direction and covering the range $-250<z<+250$~cm along the beam axis~\cite{Alme2010316}.  
It provides charged-particle trajectory reconstruction with up to 159 space points per track in the pseudorapidity range $|\eta|<0.9$ and in the full azimuth. 
The primary interaction vertex position and covariance matrix are determined from tracks reconstructed from hits in the TPC and the ITS via a $\chi^2$ analytic minimisation method.  
 
The TOF detector is equipped with Multi-gap Resistive Plate Chambers (MRPCs)~\cite{Abelev:2014ffa}. It is placed at radii between 377 cm and 399 cm, and has the same pseudorapidity and azimuthal coverage as the TPC.  The TOF measures the flight times of charged particles from the interaction point to the detector with an overall resolution of about 85 ps. For events with the 20\% lowest multiplicities, the resolution decreases to about 120 ps due to a worse start-time (collision-time) resolution. The start-time of the event is determined by combining the time estimated using the particle arrival times at the TOF and the time measured by the T0 detector, an array of Cherenkov counters located at +350 cm and -70 cm along the beamline.  
Particle identification (PID) is performed by comparing the measurement of the specific energy deposition ${\rm d}E/{\rm d}x$ in the TPC and the time-of-flight information from the TOF with the respective expected values for each mass hypothesis.

The V0 detector consists of two arrays of scintillator tiles covering the pseudorapidity regions $-3.7 < \eta < -1.7$ (V0C) and $2.8 < \eta < 5.1$ (V0A)~\cite{Abbas:2013taa}.  
The data sample analysed in this paper was collected with a minimum-bias interaction trigger requiring at least one hit in both V0A and V0C counters coincident with the arrival time of the proton and lead bunches.  
The ZDC is composed of two sets of neutron (ZNA and ZNC) and proton (ZPA and ZPC) calorimeters positioned on either side of the interaction point at $z = \pm 112.5$~m.  
Contamination from beam-background interactions was removed via offline selections based on the timing information provided by the V0 and the ZNA.  
The signals registered by the SPD and V0 detectors were used to determine the event charged-particle multiplicity; the SPD, V0 and ZDC detectors were also exploited to classify the events in centrality classes, as will be described in Sec.~\ref{sec:AnalysisMethods}.  
 
The data sample used in this paper was recorded in January 2013, during the \pPb~LHC run.  
Protons with an energy of 4 TeV were collided with Pb ions with an energy of 1.58 TeV per nucleon, resulting in collisions at a centre-of-mass energy per nucleon pair, $\sqrtsNN$, of $5.02$~TeV. 
With this beam configuration, the  centre-of-mass system moves with a rapidity of $\Delta y_{\rm cms} = 0.465$ in the direction of the proton beam, due to the different energies per nucleon of the proton and the lead beams.  
In the case of the D-meson analyses presented here, performed in the laboratory reference interval $|y_{\rm lab}|<0.5$, this leads to a shifted centre-of-mass rapidity coverage of $-0.96 < y_{\rm cms} < 0.04$.  
In the following, we will use the notation $\eta$ and $y_{\rm lab}$ to refer to the pseudorapidity and rapidity values in the laboratory reference frame, and $\eta_{\rm cms}$ and $y_{\rm cms}$ for the values evaluated in the centre-of-mass reference frame. 
A total of $10^{8}$ minimum-bias triggered events, corresponding to an integrated luminosity of $\mathcal{L}_{\rm int} = 48.6 \pm1.6~\mu$b$^{-1}$,  passed the selection criteria and were analysed.

\section{Multiplicity determination}
\label{sec:AnalysisMethods}

The production of D mesons in \pPb~collisions has been studied as a function of charged-particle multiplicity using two different observables. 
 
One observable is the $\pt$-differential nuclear modification factor,  
which is defined as the ratio of the $\pt$-differential yields measured  
in \pPb~collisions in centrality intervals to those in \pp~collisions, scaled  
by the number of binary nucleon--nucleon collisions. 
The centrality intervals were defined using three different estimators based on the multiplicity in the SPD and V0A detectors and the energy  
deposited in the zero-degree neutron calorimeter in the Pb-going side (ZNA).  
The procedure used to determine the number of binary nucleon-nucleon collisions for each event class is described in Sec.~\ref{sec:DefEventActivity} and~\cite{Adam:2014qja}.  
 
The other observable, referred to as the relative yield, is defined as the ratio of the per-event D-meson yields in \pPb~collisions in different multiplicity intervals normalised to the multiplicity-integrated yields.  
Details on the evaluation of the charged-particle multiplicity are discussed in Sec.~\ref{sec:DefRelativeMult}. 
In this analysis, the values of multiplicity measured in two different pseudorapidity intervals, namely at mid-rapidity with the SPD and at large rapidity in the Pb-going direction with the V0A, were considered. 

\subsection{Centrality estimators and $\TpPb$ determination}
\label{sec:DefEventActivity}

A centrality-dependent measurement of the nuclear modification factor requires 
the  \pPb~data sample to be sliced into classes according to an experimental observable related to the collision  
centrality, as well as a determination of the average nuclear overlap function $\langle \TpPb \rangle$, 
which is proportional to the number of nucleon--nucleon collisions $\Ncoll$, for each centrality class. 
 
The minimum-bias \pPb~data sample was divided into four centrality classes by exploiting the information from: 
{(i)} \VzA, the amplitude of the signal measured by the V0 scintillator array located in the Pb-going side, covering $2.8<\eta<5.1$, which is proportional to the number of charged particles produced in this pseudorapidity interval;   
{(ii)} \CL, the number of clusters in the outer layer of the SPD, covering $|\eta|<1.4$, which is proportional to the number of charged particles at mid-rapidity; and  
{(iii)} \ZNA, the energy deposited in the Zero Degree Neutron Calorimeter positioned in the Pb-going side by the slow nucleons produced in the interaction by nuclear de-excitation processes, or  
knocked out by wounded nucleons. 
The multiplicity of these neutrons is expected to grow monotonically with the number of binary collisions, $\Ncoll$.

Centrality classes were defined as percentiles of the visible cross section, which was measured to be ($2.09 \pm 0.07$)~b~\cite{Abelev:2014epa}.  
For the centrality classes defined using the \CL~and \VzA~multiplicities, a Glauber Monte Carlo was used to calculate the relevant geometrical quantities, namely the average numbers of participant nucleons  
$\langle \Npartglauber \rangle$, of binary collisions $\langle \Ncollglauber \rangle$, and the average nuclear overlap function $\langle \TpPbglauber \rangle$~\cite{Adam:2014qja}.  
For the case where the \ZNA~information was used, the values of $\Npart$, $\Ncoll$ and $\TpPb$ were  
obtained using the so-called hybrid method~\cite{Adam:2014qja}.  
In this approach, the determination of $\langle \TpPb \rangle$ in a given \ZNA-energy class relies on 
the assumption that the charged-particle multiplicity measured at mid-rapidity ($-1<\eta_{\rm cms}<0$)  
scales with the number of participant nucleons, $\Npart$.  
\begin{equation} 
\langle \Ncollmult \rangle_i =  \langle \Npartmult \rangle_i - 1 =  \langle \Npart^{\rm MB} \rangle \cdot  \bigg (\frac{\langle \dNdEta \rangle_i}{\langle \dNdEta \rangle^{\rm MB}}\bigg )_{-1<\eta<0} - 1\,, 
\qquad 
{\rm and} 
\qquad 
\langle \TpPbmult \rangle = \frac{\langle \Ncollmult \rangle_i}{\sigma_{\rm NN}}\,, 
\end{equation} 
where $\langle \Npart^{\rm MB} \rangle=7.9$ is the average number of participants in minimum-bias collisions and $\sigma_{\rm NN}=(70\pm5)~\mb$ is the interpolated inelastic nucleon-nucleon cross section at $\sqrtsNN=5.02~\tev$~\cite{Adam:2014qja}.  
The values of $\langle \TpPb \rangle$ obtained with the three estimators in the four multiplicity (zero-degree energy) classes used for the analysis are reported in Table~\ref{tab:ncoll}.  
 
It was demonstrated by the studies of charged-particle production reported in~\cite{Adam:2014qja} that when centrality classes are defined in \pPb~collisions, some biases are present.  
Firstly, there is a multiplicity selection bias due to the large multiplicity fluctuations for \pPb~interactions at a given impact parameter, which are comparable in magnitude to the full dynamic range of the minimum-bias multiplicity distribution.  
In addition, there is a jet-veto bias due to the contribution to the overall multiplicity from particles arising from the fragmentation of partons produced in hard-scattering processes. This causes low- (high-) multiplicity \pPb~collisions to correspond to a lower (higher) number of hard scatterings per nucleon-nucleon collision. 
Furthermore, a purely geometrical bias was suspected to affect peripheral collisions for all centrality estimators, due to the fact that the mean impact parameter between the proton and each nucleon of the Pb nucleus, calculated from a Monte Carlo Glauber simulation, rises significantly for $\Npart<6$, thus reducing the average number of multi-parton interactions for peripheral collisions.

These biases cause the nuclear modification factor of charged particles to differ from unity in the centrality classes even in the absence of nuclear effects.  
These biases decrease with increasing rapidity separation between the centrality estimator and the region where the nuclear modification factor is measured. 
A strong selection bias is observed for the \CL~estimator,  
due to the full overlap with the tracking region, which is reduced with the \VzA~estimator. 
By contrast, the selection based on the energy deposited in the \ZNA~is expected to be free from  
the biases related to the event selection, and is only affected by the geometrical bias.  
 
For these reasons, the results based on the \ZNA~selection, which is the least biased~\cite{Adam:2014qja}, provide insight into possible centrality-dependent nuclear effects on charm production in \pPb~collisions. Moreover, the measurements of the D-meson nuclear modification factor in centrality intervals defined with the three estimators described above offer the possibility to study these biases based on heavy-flavour production, which, due to the large mass of the charm quarks, is expected to scale with the number of binary collisions over the whole $\pt$ range, provided that cold nuclear matter effects are negligible.  
This is in contrast to the charged-particle yield, where a scaling with $\Ncoll$ is expected to occur only in the high-$\pt$ region.

\begin{table}[h] 
\begin{center} 
\begin{tabular}{c|ccc|cc} 
\hline 
Centrality & \multicolumn{3}{c|}{$\langle \TpPb \rangle$ Glauber-NBD ($\mb^{-1}$)}   & \multicolumn{2}{c}{$\langle \TpPb \rangle$ hybrid method ($\mb^{-1}$)}  \\ 
 (\%) &  \VzA    & \CL & Syst. (\%) & \ZNA & Syst. (\%) \\ \hline 
0--20                                                                & 0.183 & 0.190 & 11                                                  & 0.164                                                                       & 6.5                                                                         \\  
20--40                                                               & 0.134  & 0.136  & 3.7                                                 & 0.136                                                                      & 3.9                                                                         \\  
40--60                                                               & 0.092 & 0.088  & 5.0                                                   & 0.101                                                                      & 5.9                                                                         \\  
60--100                                                              & 0.037 & 0.037 & 23                                                  & 0.046                                                                       & 6.2                                                                         \\ \hline 
\end{tabular} 
\caption{$\langle \TpPb \rangle$ values in \pPb~collisions at $\sqrtsNN=5.02$~TeV obtained with a Glauber-model based approach for \VzA~and \CL, and from the hybrid method for \ZNA, as described in~\cite{Adam:2014qja}.} 
\label{tab:ncoll} 
\end{center} 
\end{table}

\subsection{Relative event multiplicity determination}
\label{sec:DefRelativeMult}

The charged-particle multiplicity, $\Nch$, was estimated at mid-rapidity by measuring the number of tracklets, $\Ntrk$, reconstructed in the SPD. 
A tracklet is defined as a track segment that joins a pair of space points on the two SPD layers and is aligned with the reconstructed primary vertex. $\Ntrk$ was counted within $|\eta| < 1.0$. 
 
The pseudorapidity acceptance of the SPD depends on the position of the interaction vertex along the beam line $\zvtx$, both due to the asymmetry of the collision system and the limited coverage of the detector. In addition, the overall SPD acceptance varies as a function of time due to a varying number of active channels.  
A data-driven correction was applied to the $\Ntrk$ distributions on an event-by-event basis to account for these two effects.  
This was done by renormalising the $\Ntrk$ distributions to the overall minimum with a Poissonian smearing to account for the fluctuations. 
Multiplicity classes were then defined based on the percentiles of analysed events in each $\Ntrk$ range.   
 
 
The conversion of $\Ntrk$ to $\Nch$ was performed using minimum-bias Monte Carlo simulations. The distribution of the measured $\Ntrk$ as a function of the number of generated ``physical primaries'' ($\Nch$) in the simulation was considered for this purpose. Physical primaries are defined as prompt particles produced in the collision and their decay products, excluding those from weak decays of strange particles.  
The proportionality factor was evaluated from a linear fit to the distribution, and was then applied to the mean $\Ntrk$ in each interval to give the estimated $\Nch$ values.  
These values were then divided by the width of the considered $\eta$ range, $\Delta\eta=2$, to give an estimated $\dNdEta$. 
The uncertainty of the $\Ntrk$ to $\Nch$ conversion was estimated by testing its deviation from linearity.  
A linear fit to the distribution was performed in each multiplicity interval to evaluate the possible changing slope of the distribution between intervals. From these fits, a series of scaling factors were obtained and compared to the multiplicity-integrated one, resulting in a 5\% uncertainty.  
 
The results are given as a function of the relative charged-particle multiplicity, $(\dNdEta)/\av{\dNdEta}$, where $\av{\dNdEta}=17.64 \pm 0.01 \, (\mathrm{stat.}) \pm 0.68 \, (\mathrm{syst.})$ was measured by ALICE for inelastic p--Pb collisions at $\sqrtsNN=5.02$~TeV with at least one charged particle within $|\eta| < 1.0$ \cite{ALICEpPbmult}. 
The $\Ntrk$ ranges considered in this analysis, and the corresponding relative multiplicity values, are given in Table~\ref{tab:multrangesSPD}.

The production of D mesons was also studied as a function of charged-particle multiplicity in the region $2.8<\eta<5.1$, as measured with the signal amplitude in the V0A detector, $\Nvzero$, reported in units of the minimum-ionising-particle charge. This estimator allows the multiplicity and the D-meson yields to be evaluated in two different pseudorapidity intervals (backward and central $\eta$), avoiding possible auto-correlations. 
 
The average $\Nvzero$ depends on $\zvtx$, due to the varying distance between the primary vertex and the detector array. This effect was corrected with the same method used for the $\Ntrk$ case, leading to an overall average $\Nvzero$ of 82.7. In this case, the results are considered as a function of the V0A multiplicity relative to the mean multiplicity in the same rapidity region, rather than performing a conversion to $\dNdEta$. The $\Nvzero$ intervals considered, and the corresponding relative multiplicity intervals, are reported in Table~\ref{tab:multrangesvzero}. 
 
It should be noted that the analyses performed as a function of centrality examine the events in samples populated by 20\% of the analysed events (40\% for the most peripheral events, see Table~\ref{tab:ncoll}), whereas those performed as a function of charged-particle multiplicity explore events from low to extremely high multiplicities, corresponding to about 60\% and 5\% of the analysed events, respectively (see Tables~\ref{tab:multrangesSPD} and~\ref{tab:multrangesvzero}). For the latter analyses, the event classes were defined to study the D-meson yield at extreme multiplicities.

\begin{table}[h]\centering  
\begin{tabular}{c c c c c} 
\hline 
$\Ntrk$ & $\dNdEta$ & $(\dNdEta)/\av{\dNdEta}$& $N^{\Dzero}_\mathrm{events} / 10^6$\\ \hline 
$[1,21]$   &  \phantom{0}9.8	 & 0.56  		& 59.0\phantom{0} \\ 
$[22,28]$ & 	23.9			 & 1.36		 & 12.8\phantom{0} \\ 
$[29,34]$ &	 30.3			& 1.72		 & \phantom{0}8.0\phantom{0} \\ 
$[35,43]$ &	 37.3			& 2.11		& \phantom{0}7.6\phantom{0} \\ 
$[44,69]$ &	 50.3			& 2.85		& \phantom{0}6.4\phantom{0} \\ 
$[70,199]$ & 	75.3	 		& 4.27		&  \phantom{0}0.47 \\   \hline 
 
\end{tabular}  
\caption{ 
\label{tab:multrangesSPD} 
Summary of the multiplicity intervals at central rapidity used for the analyses. The number of reconstructed tracklets $\Ntrk$, the average charged-particle multiplicity $\dNdEta$ (uncertainty of 5\% not quoted), and the relative charged-particle multiplicity $(\dNdEta)/\av{\dNdEta}$ (uncertainty of 6.3\% not quoted) are listed (see Sec.~\ref{sec:ntrksyst} for the uncertainties description). The number of events analysed for the $\Dzero$-meson analysis is also reported for each multiplicity range.  
} 
\end{table}

 \begin{table}[h]\centering  
 
\begin{tabular}{c c c c c}  
\hline 
$\Nvzero$ 		& $(\Nvzero)/\av{\Nvzero}$& $N^{\Dzero}_\mathrm{events} / 10^6$\\ \hline 
$[0,90]$   	& 	 		0.48		&  60.3 \\ 
$[91,132]$ 	& 		1.32		& 15.3 \\ 
$[133,172]$ 	& 		1.81		&  \phantom{0}9.7 \\ 
$[173,226]$	& 		2.36		& \phantom{0}6.5 \\ 
$[227,798]$ 	& 		3.29		& \phantom{0}4.0 \\\hline  
\rule{0pt}{2.3ex}$[173,798]$	& 		2.72		& 10.5 \\\hline 
 
\end{tabular}  
\caption{\label{tab:multrangesvzero} 
Summary of the multiplicity intervals at backward rapidity used for the analyses. The V0A signal $\Nvzero$ intervals and the relative multiplicity $(\Nvzero)/\av{\Nvzero}$ (uncertainty of 5\% not quoted) are listed (see Sec.~\ref{sec:ntrksyst} for the uncertainties description). The number of events analysed for the $\Dzero$-meson analysis is also reported for each multiplicity range. 
} 
\end{table}

\FloatBarrier
\section{D meson reconstruction}
\label{sec:Dreco}

The $\Dzero$, $\Dplus$, and $\Dstar$ mesons were reconstructed via their hadronic decay channels $\DtoKpi$ (with a branching ratio, BR, of $3.88\pm 0.05\%$), $\DtoKpipi$ (BR of $9.13\pm0.19\%$), and $\DstartoDpi$ (BR of $67.7\pm 0.05\%$) followed by $\DtoKpi$, and their corresponding charge conjugates~\citep{PDG}. The $\Dzero$ and $\Dplus$ weak decays, with mean proper decay lengths ($c\tau$) of about 123 and 312~$\mu$m, respectively, were selected from reconstructed secondary vertices separated by a few hundred microns from the interaction point. The $\Dstar$ meson decays strongly at the primary vertex, and the decay topology of the produced $\Dzero$ was reconstructed along with a soft pion originating at the primary vertex. 
 
Events were selected by requiring a primary vertex within $\pm 10$~cm from the centre of the detector along the beamline. An algorithm to detect multiple interaction vertices was used to reduce the pile-up contribution.  
 $\Dzero$ and $\Dplus$ candidates were defined using pairs or triplets of tracks with the proper charge sign combination, within the fiducial acceptance $|\eta|<0.8$ and with transverse momentum $\pt>0.3$~GeV/$c$. 
Only good quality tracks were considered in the combinatorics by requiring selection criteria as described in~\cite{ALICE:2011aa,Abelev:2012vra,Abelev:2014hha}.   
The selection of tracks with $|\eta|<0.8$ reduces the D-meson acceptance, which drops steeply to zero for $|\ylab|>0.5$ at low $\pt$ and for $|\ylab|>0.8$ at $\pt>5$~GeV/$c$.  
Therefore, a $\pt$-dependent fiducial acceptance region was defined, as reported in~\cite{ALICE:2011aa,Abelev:2012vra,Abelev:2014hha}.   
 
The selection strategy of the D-meson decay topology was based on the displacement of the decay tracks from the interaction vertex, the separation between the secondary and primary vertices, and the pointing angle, defined as the angle between the reconstructed D-meson momentum and its flight line (the vector between the primary and  the secondary vertices). The cuts on the selection variables were chosen in order to obtain a large statistical significance of the D-meson signals, as well as an as large as possible selection efficiency. Therefore, the cut values depend on the D-meson $\pt$ and species.  
In the case of the analysis of the relative yields as a function of multiplicity, the same selections were used in all multiplicity intervals in order to minimise the effect of the efficiency corrections on the ratio of the yields in the multiplicity intervals to the multiplicity-integrated ones. 
On the other hand, for the analysis of the nuclear modification factor in different centrality classes, the cut values were optimised in each centrality class. 
Particle identification criteria were applied on the decay tracks, based on the TPC and TOF detector responses, in order to obtain a further reduction of the combinatorial background as explained in~\cite{ALICE:2011aa,Abelev:2012vra,Abelev:2014hha}.

The raw D-meson yields, both multiplicity-integrated and in each multiplicity or centrality class, were extracted in the considered $\pt$ intervals by means of a fit to the invariant mass ($M$) distributions of the selected candidates (for the $\Dstar$ meson the mass difference distributions $\Delta M=M({\rm K}\pi\pi)-M({\rm K}\pi)$ were used). The fit function is the sum of a Gaussian to describe the signal and a function describing the background shape, which is an exponential for $\Dzero$ and $\Dplus$ and a threshold function multiplied by an exponential ($a\sqrt{\Delta M-M_{\pi}}\cdot e^{b(\Delta M-M_{\pi})}$, where $M_{\pi}$ is the pion mass and $a$ and $b$ are free parameters) for the $\Dstar$. The centroids and the widths of the Gaussian functions were found to be in agreement with the world average D-meson masses and the values obtained in simulations, respectively, in all multiplicity, centrality and $\pt$ intervals. In particular, the widths of the Gaussian functions are independent of multiplicity (or centrality) and increase with increasing D-meson $\pt$. In the relative yield analysis, in order to reduce the effect of the statistical fluctuations, the fits were performed by fixing the Gaussian centroids to the world average D-meson masses, and the widths to the values obtained from a fit to the invariant mass distribution in minimum-bias events, where the signal statistical significance is larger.  
 
\begin{figure} 
\begin{center} 
\includegraphics[width=0.9\textwidth]{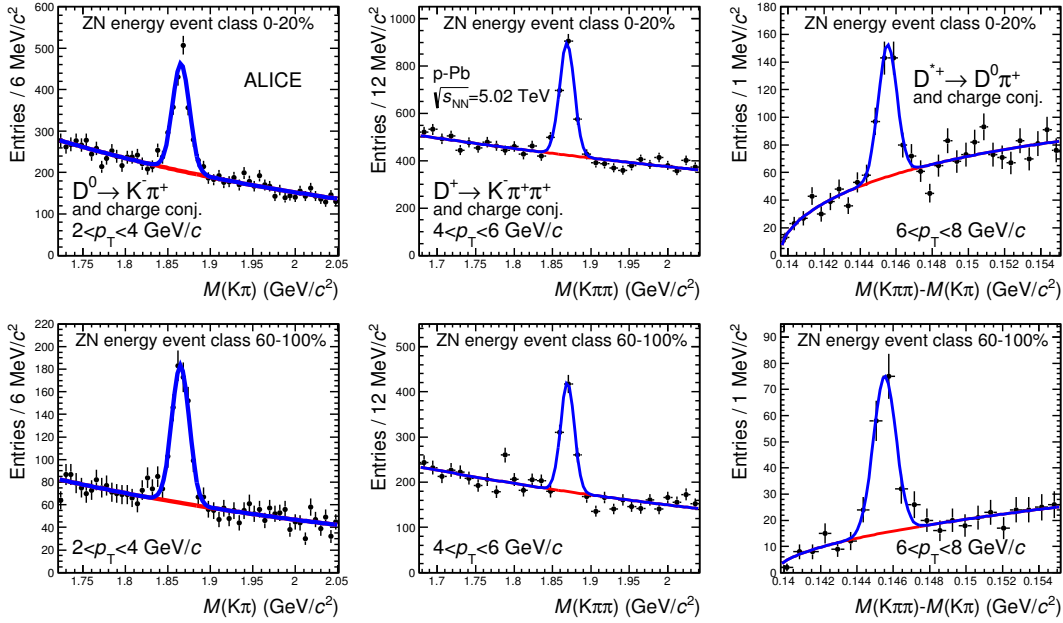} 
\caption{Distributions of the invariant mass for $\Dzero$ (left column) and $\Dplus$ (middle column) candidates and of the mass difference for $\Dstar$ candidates (right column) in two centrality classes defined with the ZNA estimator: 0--20\% and 60--100\%. The red lines in each plot represent the fit to the background, and the blue lines represent the sum of signal and background. One $\pt$ interval is shown for each meson species: $2<\pt<4$~GeV/$c$ for $\Dzero$, $4<\pt<6$~GeV/$c$ for $\Dplus$, and $6<\pt<8$~GeV/$c$ for $\Dstar$.} 
\label{fig:DmesonInvMassFits} 
\end{center} 
\end{figure} 
Figure~\ref{fig:DmesonInvMassFits} shows the $\Dzero$ and $\Dplus$ invariant mass, and $\Dstar$ mass difference distributions in the $2<\pt<4$~GeV/$c$, $4<\pt<6$~GeV/$c$, $6<\pt<8$~GeV/$c$ intervals, respectively, for the 0--20\% and 60--100\% centrality classes defined with the ZNA estimator.  
The fits to the invariant mass distributions were repeated under different conditions and the raw yields were extracted by using alternative methods in order to determine the systematic uncertainties related to the extraction of the raw D-meson counts. The fits were performed by varying the invariant mass ranges and bin widths of the histograms, and considering different functions to describe the background, namely parabolic or linear functions. The raw yields were also obtained by counting the entries of the histograms within a $3\sigma$ interval centred on the peak position, after the subtraction of the background estimated from a fit to the side bands, far away from the D-meson peaks.  

The raw counts of D mesons extracted in each $\pt$ and multiplicity interval were corrected for the acceptance and the reconstruction and selection efficiency. The correction factor for each D-meson species was obtained by using Monte Carlo simulations.  
Events containing a $\rm{c\bar{c}}$ or $\rm{b\bar{b}}$ pair were generated by using the PYTHIA v6.4.21 event generator~\cite{Sjostrand:2006za} with the Perugia-0 tune~\cite{Skands:2009zm} and adding an underlying event generated with HIJING v.1.36~\cite{Wang:1991hta}. 
Detailed descriptions of the detector response, the geometry of the apparatus and the conditions of the luminous region were included in the simulation.  
The generated D-meson $\pt$ distribution was tuned in order to reproduce the FONLL~\cite{Cacciari:1998it} spectrum at $\sqrt{s} = 5.02$~TeV.  
The reconstruction and selection efficiency depends on the multiplicity of charged particles produced in the collision, since the primary vertex resolution and the resolution on the topological selection variables improve at high multiplicity.  
The generated events were weighted on the basis of their charged-particle multiplicity in order to match the multiplicity distribution observed in the data.  
The reconstruction and selection efficiency depends on the D-meson species and on $\pt$. For prompt $\Dzero$ mesons it is about 1--2\% in the $1<\pt<2~\gevc$ interval, where the selection criteria are more stringent due to the higher combinatorial background, and it increases to 20\% in $12<\pt<24~\gevc$. The efficiency for D mesons from B decays is higher because the decay vertices of feed-down D mesons are more displaced from the primary vertex and they are more efficiently selected by the topological selections. The efficiencies are slightly larger at high multiplicity, by about 4--10\%.

The D-meson raw yields have two components: the prompt D-meson contribution (produced in the charm quark fragmentation, either directly or through strong decays of excited open charm states) and the feed-down contribution originating from B-meson decays. The yield of D mesons from B decays was subtracted from the raw counts by applying a correction factor, $\fprompt$, which represents the fraction of promptly produced D mesons.  
The $f_{\rm prompt}$ factor was evaluated using the B-hadron production cross section obtained from the FONLL pQCD calculation~\cite{Cacciari:1998it,Cacciari:2001td,Cacciari:2012ny}, the ${\rm B}\rightarrow{\rm D}+X$ kinematics from the EvtGen package~\cite{Lange2001152}, and the 
acceptance times efficiency for D mesons from B decays obtained from the Monte Carlo simulations~\cite{ALICE:2011aa}.  
The value of $\fprompt$ depends on the nuclear modification factor, $R_{\rm pPb}^{\rm feed-down}$, of the feed-down D mesons. This quantity is related to the nuclear modification of beauty production, which has not been measured in the $\pt$ interval of these analyses. Therefore, the nuclear modification factor of feed-down D mesons was assumed to be equal to that of prompt D mesons, $R_{\rm pPb}^{\rm feed-down}=R_{\rm pPb}^{\rm prompt}$, and a systematic uncertainty was assigned considering the variation $0.9<R_{\rm pPb}^{\rm feed-down}/R_{\rm pPb}^{\rm prompt}<1.3$. These assumptions were based on the study of the possible modification of the B-hadron production due to the modification of the PDFs in the nucleus through either CGC or pQCD calculations with the EPS09 parameterisation of the nPDFs~\cite{Fujii:2013yja,Eskola:2009uj}.

\section{Nuclear modification factor as a function of centrality}
\label{sec:EA}

The nuclear modification factor of prompt $\Dzero$, $\Dplus$ and $\Dstar$ mesons was studied as a function of $\pt$ using the three different centrality estimators introduced in Sec.~\ref{sec:DefEventActivity}, based on different measurements of the centrality in terms of multiplicity (\CL~and \VzA~estimators) or zero-degree energy (\ZNA~estimator). For each estimator, the analysis of D-meson production was carried out in four event classes, and the nuclear modification factor was calculated as: 
\begin{equation}  
\QpPb = \frac{(\dNdpt)^{\mathrm{cent}}_{\mathrm{pPb} }} 
{ \langle \TpPb \rangle \, \times \, (\mathrm{d} \sigma^{\rm D} / \mathrm{d} \pt )_{\mathrm{pp}}} \,, 
\label{eq:QpPb} 
\end{equation} 
where $(\dNdpt)^{\mathrm{cent}}_{\mathrm{pPb}}$ is the yield of prompt D mesons in \pPb~collisions in a given centrality class, $(\mathrm{d} \sigma^{\rm D} / \mathrm{d} \pt)_{\mathrm{pp}}$ is the cross section of prompt D mesons in \pp~collisions at the same $\sqrts$, and $\langle \TpPb \rangle$ is the average nuclear overlap function in a given centrality class, which was estimated with the Glauber-model approach for the \CL~and \VzA~estimators ($\TpPbglauber$) and with the hybrid method for the \ZNA~estimator ($\TpPbmult$) (see Sec.~\ref{sec:DefEventActivity}). 
 
In contrast to the multiplicity-integrated \mbox{$\RpPb =  (\mathrm{d} \sigma^{\rm D} / \mathrm{d} \pt )_\mathrm{pPb} / \left( A \cdot (\mathrm{d} \sigma^{\rm D} / \mathrm{d} \pt )_\mathrm{pp} \right) $}, $\QpPb$  
is influenced by potential biases in the centrality estimation that are not related to nuclear effects, as explained in Sec.~\ref{sec:DefEventActivity}. Hence, $\QpPb$ may be different from unity even in the absence of nuclear effects, in particular if measured with respect to the \CL~and \VzA~estimators. Complementary to this, the measurement of $\QpPb$ with the \ZNA~estimator allows the least biased estimation of the possible centrality-dependent modification of the $\pt$-differential yields in \pPb~collisions with respect to the binary-scaled yields in \pp~collisions. 
 
The cross sections of prompt D-meson production in \pp~collisions at $\sqrts=5.02$~TeV were obtained by a pQCD-based energy scaling of the $\pt$-differential cross sections measured at $\sqrts= 7$~TeV with the scaling factor evaluated by the ratio of the FONLL~\cite{Cacciari:1998it,Cacciari:2001td,Cacciari:2012ny} calculations at 5.02 and 7~TeV~\cite{Averbeck:2011ga}.  
The scaling procedure was validated by comparing the D-meson $\pt$-differential cross sections at 2.76 TeV with the 7 TeV data scaled down to 2.76 TeV~\cite{Abelev:2012vra}. In the case of $\Dzero$ mesons, some refinements were considered for the lowest and highest $\pt$ intervals.  
For $1< \pt < 2~\gevc$, where the $\Dzero$ cross section was measured at both 7 and 2.76~TeV~\cite{ALICE:2012ab,ALICE:2011aa}, both measurements were scaled to 5.02~TeV and averaged using the inverse squared of their relative statistical and systematic uncertainties as weights. Since the ALICE measurements of the $\Dzero$ cross section in \pp~data are limited to $\pt<16~\gevc$, the estimate for $16<\pt<24~\gevc$ was determined by extrapolating the 7 TeV cross section to higher $\pt$ using the FONLL $\pt$-differential spectrum normalised to the measurement in $5< \pt < 16~\gevc$, and scaling it down to 5.02~TeV.

\label{subSect:EAcorrection} 
 
The raw numbers of D mesons in each $\pt$ and centrality interval were extracted and corrected by the acceptance and efficiency obtained from Monte Carlo simulations, as described in Sec.~\ref{sec:Dreco}.  
The feed-down from B-hadron decays was subtracted from the extracted yields by calculating $\fprompt$ in each centrality class independently, as described in Sec.~\ref{sec:Dreco}.

\subsection{Systematic uncertainties}  
\label{subSect:EAsystematics} 
 
The systematic uncertainties (yield extraction, reconstruction and selection efficiency determination and feed-down subtraction) do not depend on the estimator used to define the centrality classes. A mild dependence of the uncertainty on the multiplicity that populates the different centrality classes was observed, resulting in slightly larger uncertainties in the event class with the lowest multiplicity.  
 
The systematic uncertainty of the yield extraction procedure was estimated by varying the fit conditions and by using the bin counting method as introduced in Sec.~\ref{sec:Dreco}. It is about 3--4\% at intermediate $\pt$ ($2< \pt < 6~\gevc$) and increases to 8--10\% at $\pt<2~\gevc$ and $\pt>6~\gevc$.  
For the $\Dzero$ meson, the yield extraction systematic uncertainty includes the contribution to the raw yield of signal candidates reconstructed by assigning the wrong mass to the final state hadrons (about 3--4\% for all $\pt$ intervals)~\cite{Abelev:2014hha}. 
 
The influence of the tracking efficiency was estimated by varying the track selection criteria. The corresponding uncertainty was found to be about 3\% per track, resulting in a total uncertainty of 6\% (9\%) for a two- (three-)particle decay.  
The uncertainty due to the D-meson candidate selection criteria was evaluated by varying the topological selections used. It was estimated to be 10\% for the interval $1< \pt<2~\gevc$ and 5\% for $\pt>2~\gevc$.  
 
The effect of the generated D-meson $\pt$ shape used to compute the efficiency was estimated by comparing the efficiency values obtained with the PYTHIA and the FONLL $\pt$ spectra. A systematic uncertainty of 2--3\% was applied only in the interval $1< \pt<2~\gevc$ due to this.  
The uncertainty due to the multiplicity dependence of the reconstruction and selection efficiency was evaluated changing the weight functions used to reproduce the measured charged-particle multiplicity in the simulations. The multiplicity weights were determined by the ratio of the distribution of the number of tracklets within $|\eta|<1$ in data and Monte Carlo. The weights were computed for: {(i)} all events selected in the analysis, {(ii)} events with a D-meson candidate within approximately $\pm10\sigma$ of the invariant mass peak, and {(iii)} events with a D-meson candidate in the $\pm3\sigma$ invariant mass region. A deviation of about 10\% is observed for D mesons at low $\pt$. For high-$\pt$ D mesons ($\pt > 12~\gevc$), the weights have a smaller effect on the efficiency determination, introducing a difference of only 4\%.

The analysis was repeated without applying the particle identification selections to the D-meson decay hadrons. The corrected yields were consistent, within statistical fluctuations, with those calculated considering particle identification selections. Therefore, no corresponding uncertainty was assigned.  
 
The systematic uncertainty due to the subtraction of feed-down D mesons from B decays was estimated by considering the FONLL uncertainties on the normalisation and factorisation scales and using a second subtraction method based on the ratio of FONLL calculations for D- and B-meson cross sections~\cite{ALICE:2011aa}. The magnitude of this systematic uncertainty depends on the meson species and on the $\pt$ interval considered in the measurement, since it is related to the topological selections applied in each analysis. As explained in Sec.~\ref{sec:Dreco}, a variation of the feed-down D-meson nuclear modification factor was also taken into account as part of the systematics. The quadratic sum of the two contributions to the $\QpPb$ was found to range from a few percent up to 30\%. 
 
The denominator of the $\QpPb$ has an uncertainty on the $\langle \TpPb \rangle$, which is reported in Table~\ref{tab:ncoll}, and an uncertainty on the \pp~reference.  
The latter has a contribution coming from the 7 TeV measurement (ranging from 15\% up to 25\%) and one from the scaling factor ranging from $^{+17\%}_{-4\%}$ at $\pt=1~\gevc$ to $\pm3\%$ for $\pt>8~\gevc$. The uncertainty on the energy scaling factor was estimated by varying the calculation parameters as described in~\cite{Averbeck:2011ga}. A larger uncertainty for $\Dzero$ in $16<\pt < 24~\gevc$ was quantified due to the extrapolation procedure explained above; in that case the uncertainty is $^{+17.5\%}_{-4\%}$.  
The global $\QpPb$ uncertainties were determined by adding the \pp~and \pPb~uncertainties in quadrature, except for the branching ratio uncertainty, which cancels out in the ratio, and the feed-down contribution, which partially cancels out.

\subsection{Results}
\label{sec:EAresultsZNA}

\begin{figure}[!htbp] 
\begin{center} 
  \includegraphics[width=0.475\columnwidth]{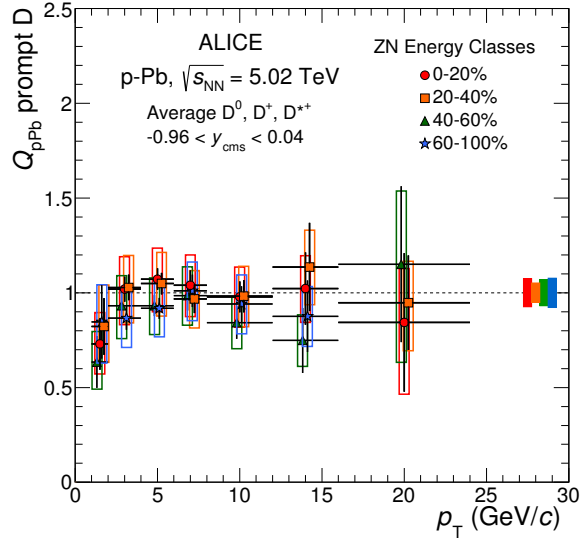} 
\caption{Average $\Dzero$, $\Dplus$ and $\Dstar$ meson nuclear 
  modification factors as a function of $\pt$ in the 0--20\%, 20--40\%, 
  40--60\% and 60--100\% centrality classes selected with the \ZNA~estimator.  
The vertical error bars and the empty boxes represent the statistical and systematic uncertainties, respectively. The colour-filled boxes at $\QpPb=1$ represent the normalisation uncertainties.  
Symbols are displaced from the bin centre for clarity.  
\label{fig:DQpPbAvZNA} 
} 
\end{center} 
\end{figure}

The nuclear modification factors of $\Dzero$, $\Dplus$ and $\Dstar$ 
mesons were calculated according to Eq.~(\ref{eq:QpPb}) in four 
centrality classes (0--20\%, 20--40\%, 40--60\% and 60--100\%) defined with the \ZNA~estimator, and applying the 
hybrid method to obtain the $\langle\TpPb\rangle$ in each class.  
\begin{figure}[!t] 
\begin{center} 
\subfigure[\ZNA~estimator, 0--20\%]{ 
	\label{fig:DQpPb_ZNA_Cmp_020} 
	\includegraphics[width=0.475\columnwidth]{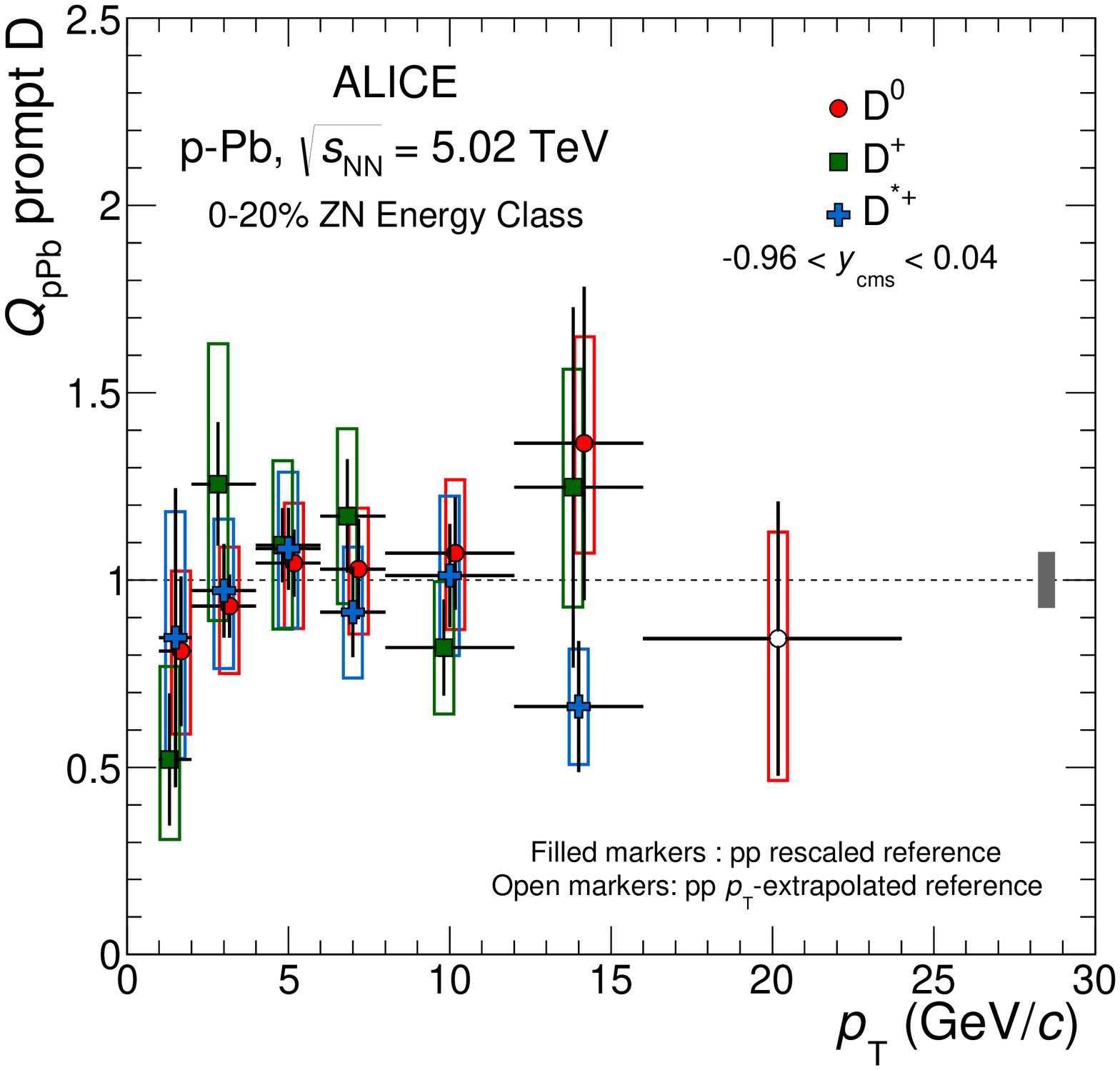} 
	}  
\subfigure[\ZNA~estimator, 40--60\%]{ 
	\label{fig:DQpPb_ZNA_Cmp_60100} 
	\includegraphics[width=0.475\columnwidth]{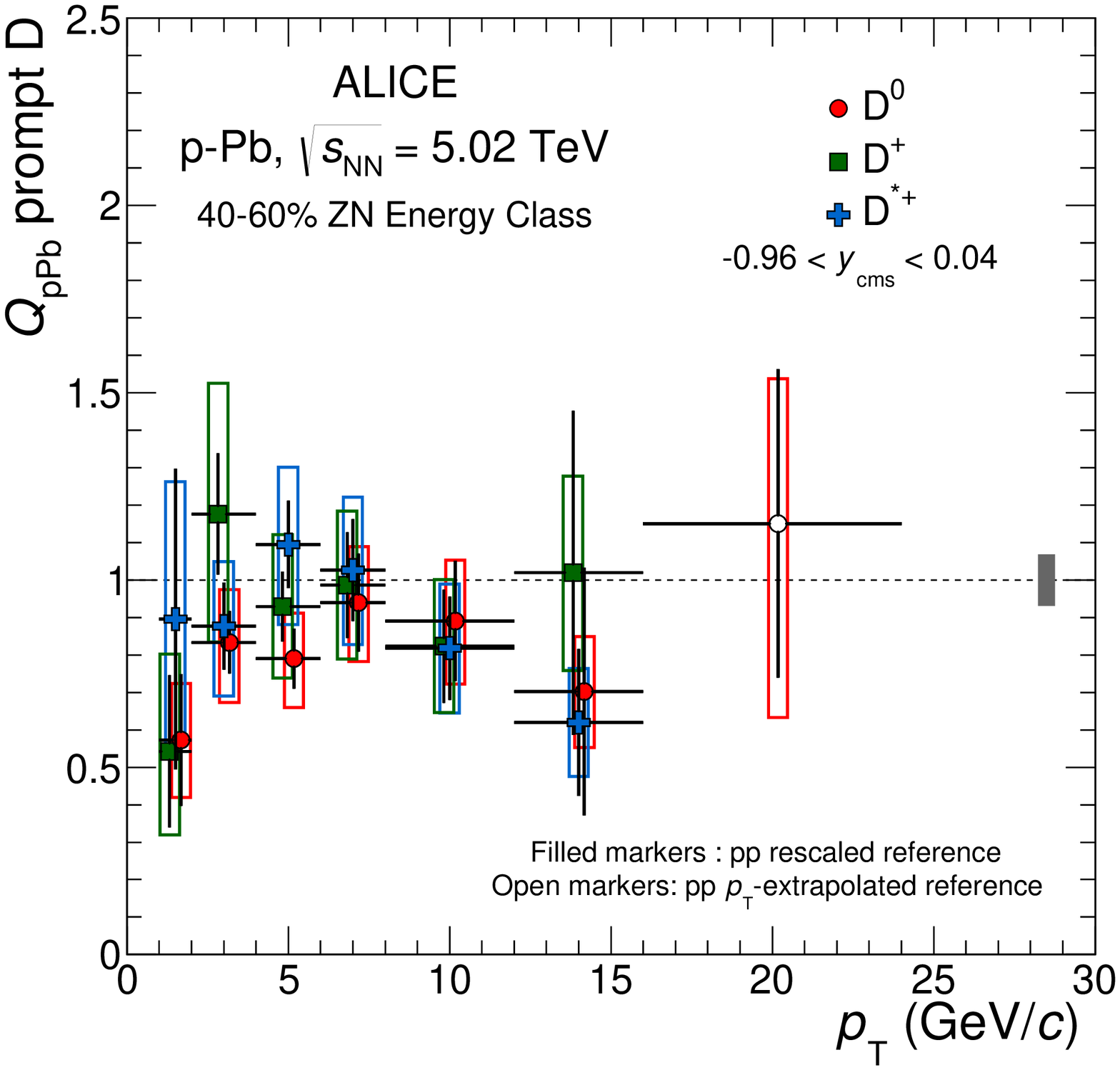} 
} 
\caption{$\Dzero$, $\Dplus$ and $\Dstar$ meson nuclear modification 
  factors as a function of $\pt$ for: (a) the 0--20\% centrality class and (b) the 
  40--60\% centrality class selected with the \ZNA~estimator. 
The vertical error bars and the empty boxes represent the statistical and systematic uncertainties, respectively.  
The grey-filled box at $Q_{\mathrm{pPb}}=1$ represents the normalisation uncertainty.  
Symbols are displaced from the bin centre for clarity.  
\label{fig:D0DpDsQpPbZNA} 
} 
\end{center} 
\end{figure} 
Figure~\ref{fig:D0DpDsQpPbZNA} illustrates these results for 0--20\% and 
40--60\% centrality classes.  
The $\QpPb$ of the three D-meson species were found to be consistent with one another within the statistical and systematic uncertainties for each $\pt$ and centrality class considered.  
Therefore, the average of the $\Dzero$, $\Dplus$ and $\Dstar$ meson 
results was evaluated in each centrality class considering the inverse square of the relative statistical uncertainties as weights.  
The systematic uncertainties on the averages were computed considering the tracking efficiency, the B feed-down subtraction and the scaling of the pp reference as correlated uncertainty sources among the three mesons.   
The averages of the $\Dzero$, $\Dplus$ and $\Dstar$ $\pt$-differential 
nuclear modification factors in different centrality classes obtained with 
the \ZNA~estimator are presented in Fig.~\ref{fig:DQpPbAvZNA} and Table~\ref{tab:DQpPbZNA}.  
The D-meson $\QpPb$ results in the different centrality classes are consistent with unity within the uncertainties in the measurement $\pt$ interval. Typical values of the $\QpPb$ uncertainties are of 7\%\,(stat.) and 16\%\,(syst.) for $2<\pt<4~\gevc$.  
It should be noted that with this centrality estimator no bias is expected due to the event selection, and only a small bias in peripheral events, due to the geometrical bias in the determination of the number of hard scatterings, was observed in the studies with charged particles~\cite{Adam:2014qja}. 
Therefore, with the least biased centrality estimator, the D-meson $\QpPb$ results are consistent within statistical and systematic uncertainties with binary collision scaling of the yield in \pp~collisions, independent of the geometry of the collision. 

\subsubsection{$\QpPb$ with $\CL$ and $\VzA$ estimators}
\label{sec:EAresultsCL1V0A}

\begin{figure}[!htbp] 
\begin{center} 
\subfigure[\CL~estimator]{ 
	\label{fig:DQpPb_CL1} 
	\includegraphics[width=0.475\columnwidth]{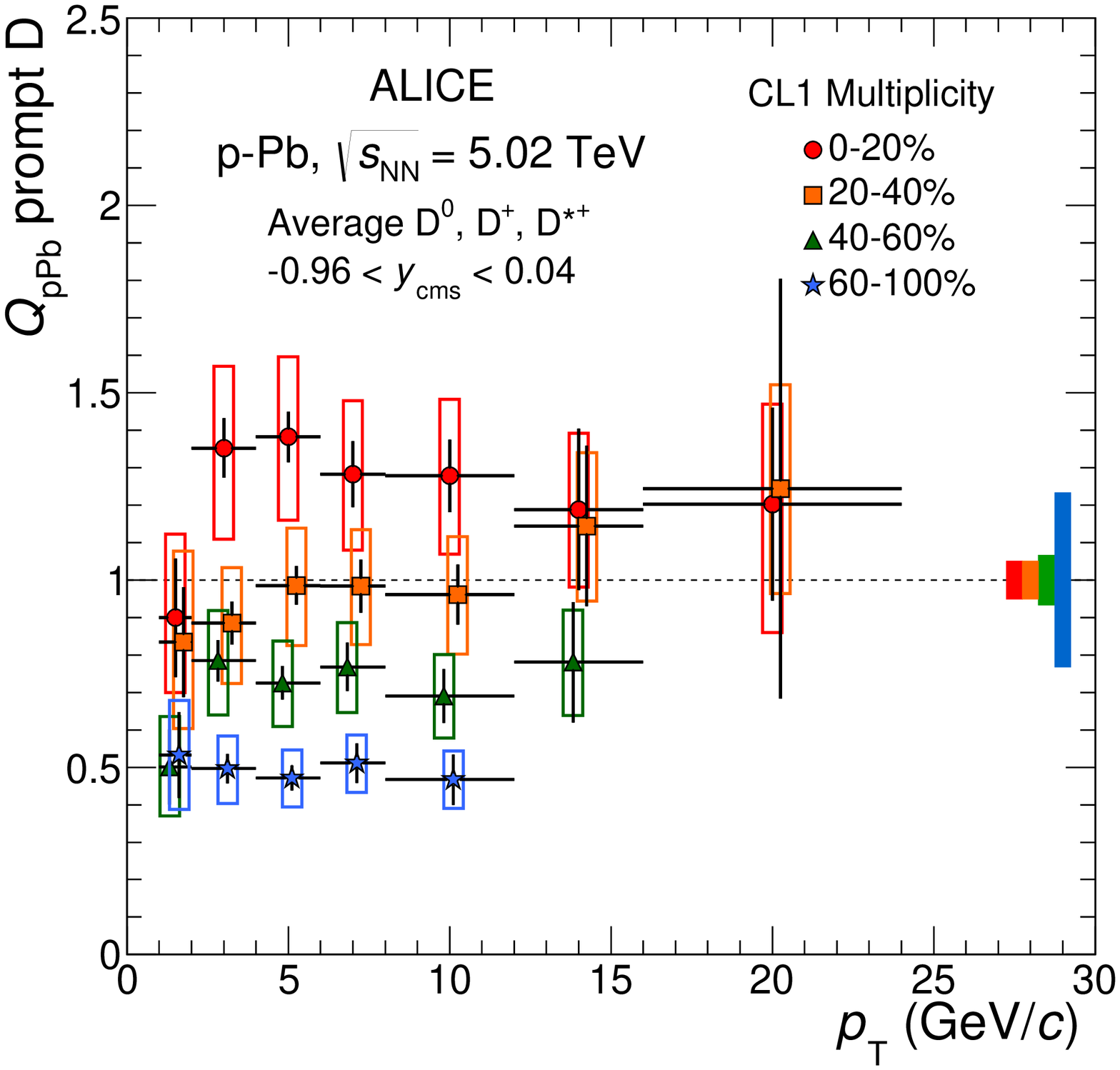} 
	}  
\subfigure[\VzA~estimator]{ 
	\label{fig:DQpPb_V0A} 
	\includegraphics[width=0.475\columnwidth]{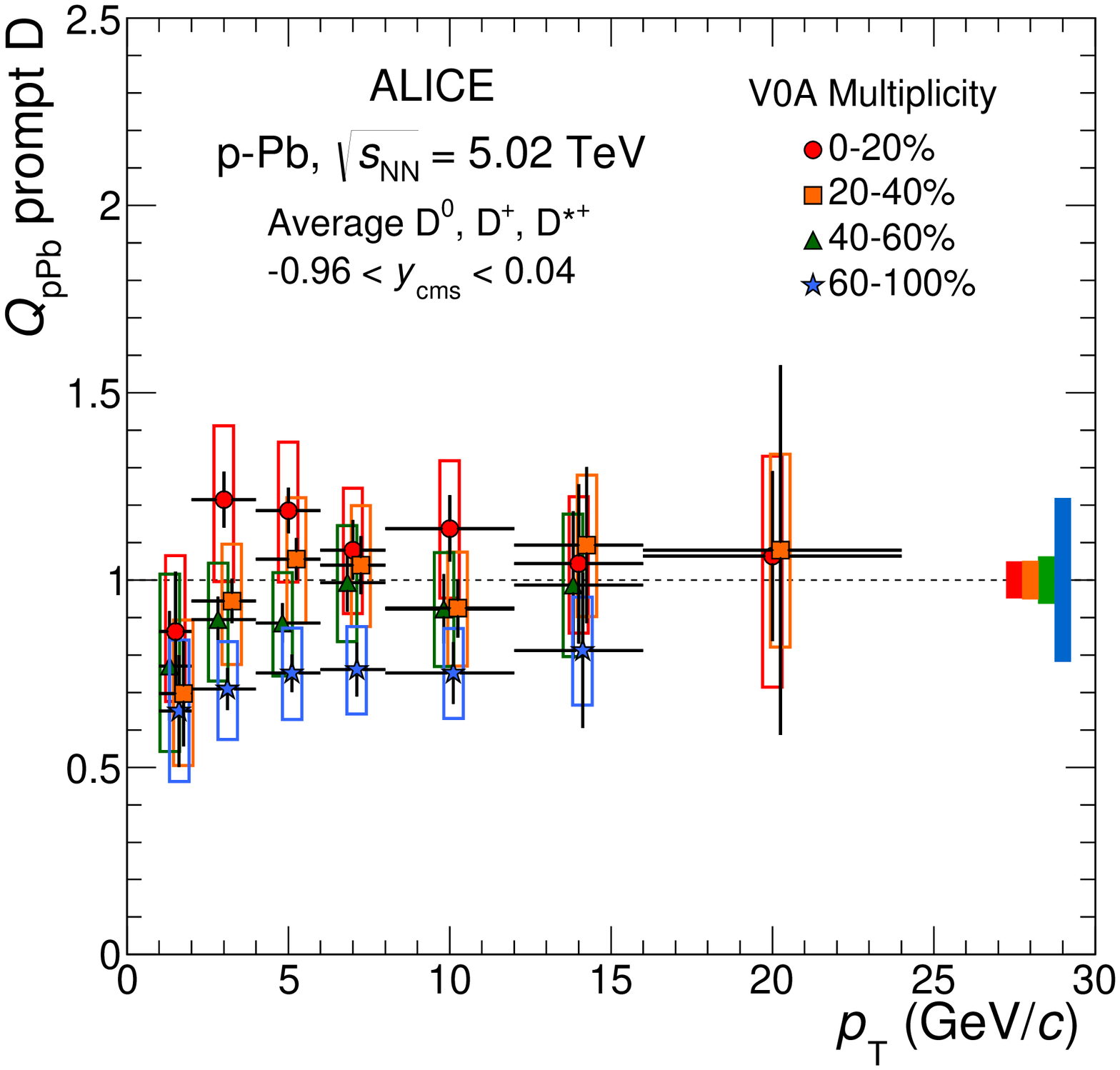} 
} 
\caption{Average $\Dzero$, $\Dplus$ and $\Dstar$ meson nuclear 
  modification factors as a function of $\pt$ in the 0--20\%, 20--40\%, 
  40--60\% and 60--100\% centrality classes selected with: (a) the 
  \CL~estimator, and (b) the \VzA~estimator.  
The vertical error bars and the empty boxes represent the statistical and systematic uncertainties, respectively. The colour-filled boxes at $\QpPb=1$ represent the normalisation uncertainties.  
Symbols are displaced from the bin centre for clarity.  
\label{fig:DQpPbAv} 
} 
\end{center} 
\end{figure}

As explained in Sec.~\ref{sec:DefEventActivity}, the  $\Dzero$, $\Dplus$ and $\Dstar$ $\QpPb$ were also calculated with the~\CL~and~\VzA~estimators in four centrality classes to study the centrality selection biases based on heavy-flavour production from low to high $\pt$.  
The $\QpPb$ results for the three D-meson species were found to be consistent with one another within the statistical and systematic uncertainties for each $\pt$ and centrality class considered.  
Therefore, the averages of the $\Dzero$, $\Dplus$ and $\Dstar$ meson 
results and the systematic uncertainties were evaluated as explained before. 
The averages of the $\pt$-differential $\Dzero$, $\Dplus$ and $\Dstar$  
nuclear modification factors in different centrality classes with 
\CL~and \VzA~estimators are presented in Fig.~\ref{fig:DQpPbAv} (see also Tables~\ref{tab:DQpPbCL1} and~\ref{tab:DQpPbV0A}).  
 
The centrality estimation from the \CL~multiplicity suffers from a large bias introduced by multiplicity fluctuations in the central rapidity region caused by fluctuations of the number of hard scatterings per nucleon collision, which affect the $\langle \TpPb \rangle$ determination~\cite{Adam:2014qja}.  
The $Q^{\mathrm{CL1}}_{\mathrm{pPb}}$ results show an ordering from low (60--100\%) to high (0--20\%) multiplicity,  
with a difference larger than a factor of two between the most central and most peripheral classes, induced by the bias on the centrality estimator. 
 
The \VzA~estimator classifies the events as a function of the multiplicity in the backward rapidity region. The rapidity gap with respect to the central rapidity D-meson analyses removes part of the event selection bias.   
The $Q^{\mathrm{V0A}}_{\mathrm{pPb}}$ values evolve from higher ($>1$) to lower ($<1$) values from the 0--20\% to the 60--100\% centrality class.  
The $Q^{\mathrm{V0A}}_{\mathrm{pPb}}$ results present a similar qualitative behaviour to the $Q^{\mathrm{CL1}}_{\mathrm{pPb}}$ ones, with a smaller difference between centrality classes.  
This is consistent with the expectation of a smaller bias when there is a rapidity gap between the regions where the centrality and the D-meson yield are studied. 

\subsubsection{Comparison with charged-particle $\QpPb$}
\label{sec:EAresultsCmpChargedParticles}

\begin{figure}[!htb] 
\begin{center} 
\subfigure[$2<\pt<4~\gevc$]{ 
	\label{fig:DQpPb3_2To4} 
	\includegraphics[width=0.475\columnwidth]{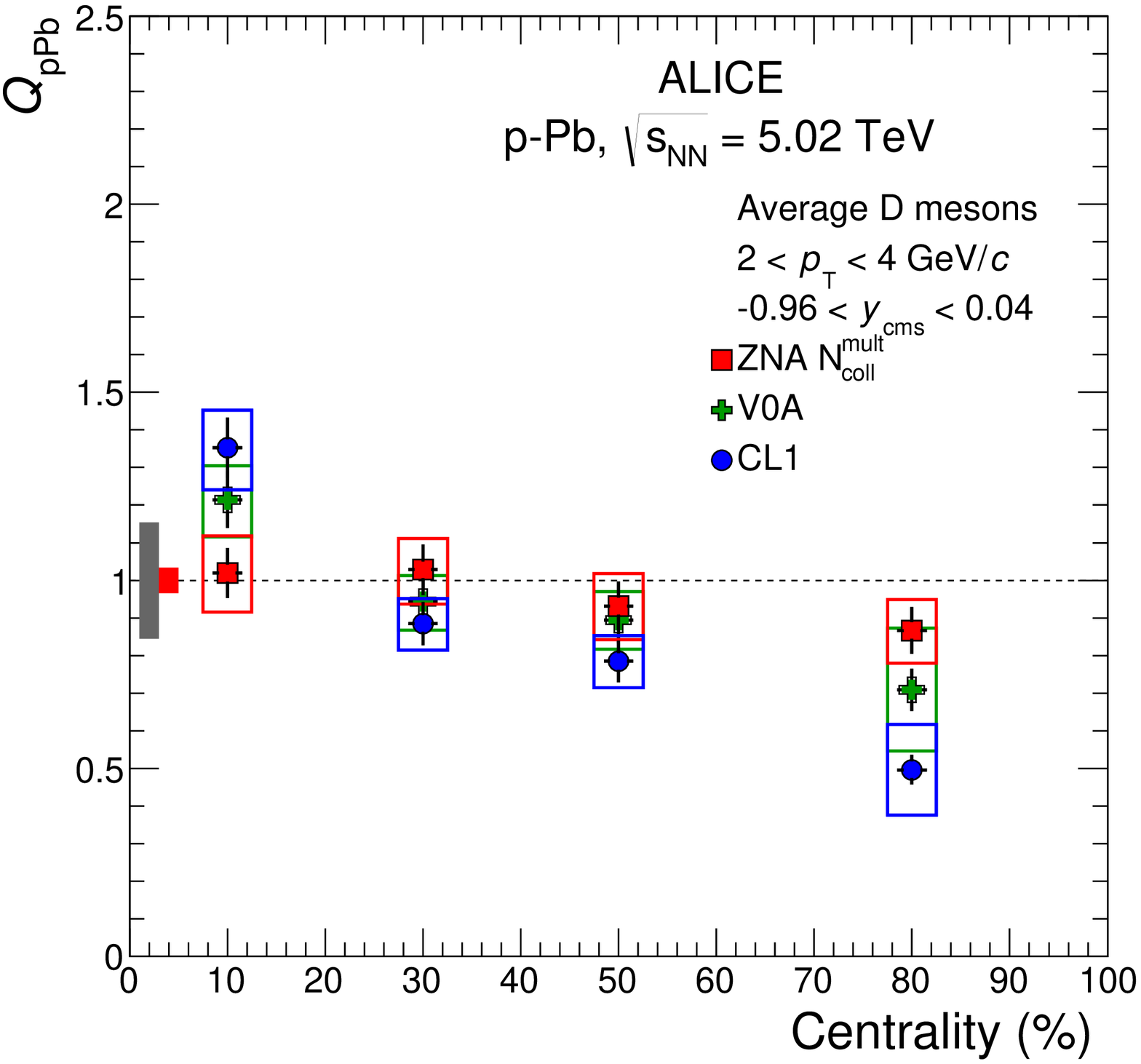} 
	}  
\subfigure[$8<\pt<12~\gevc$]{ 
	\label{fig:DQpPb3_8To12} 
	\includegraphics[width=0.475\columnwidth]{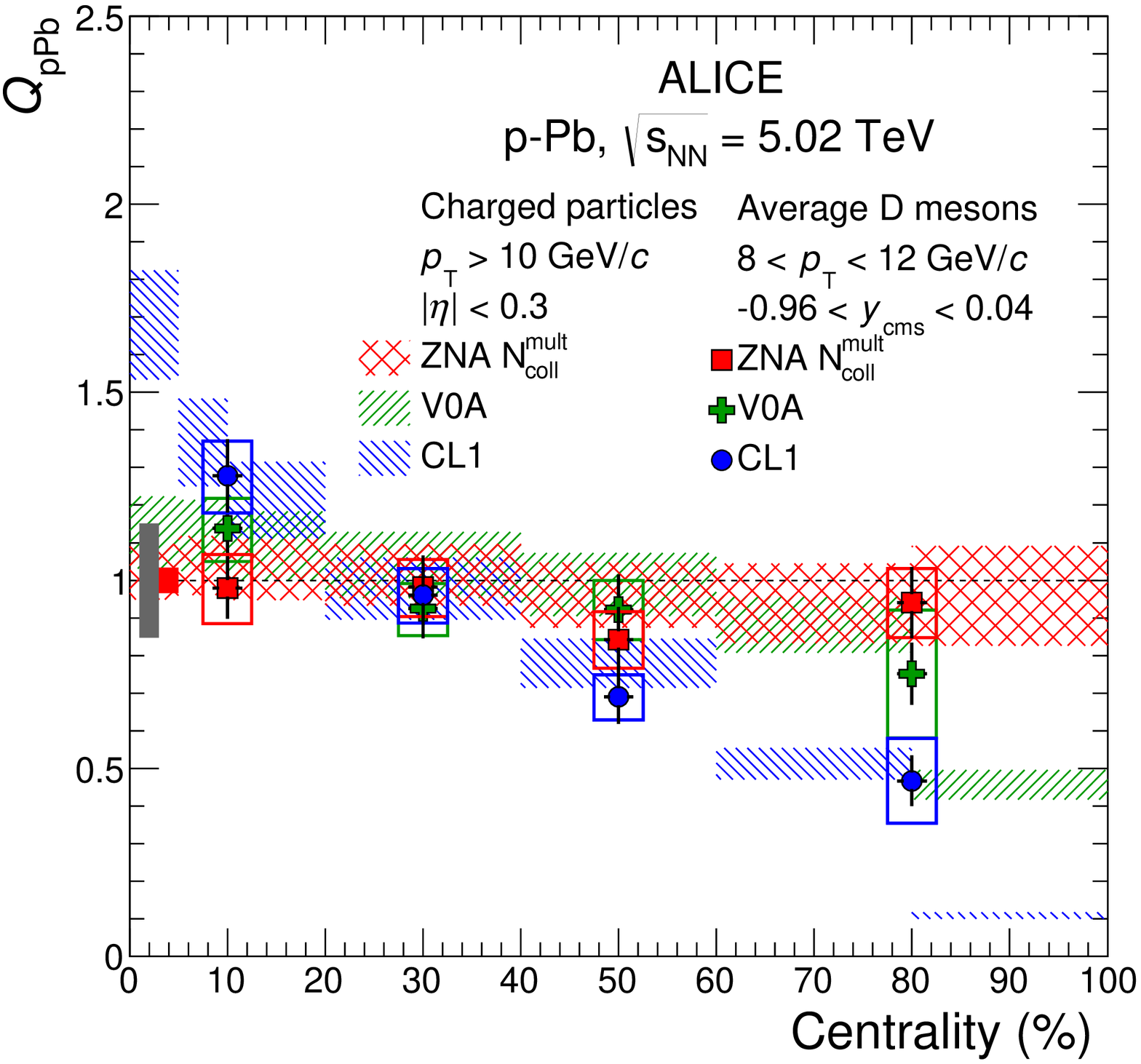} 
} 
\caption{Average $\Dzero$, $\Dplus$ and $\Dstar$ meson $\QpPb$ as a 
  function of centrality with the \CL, the \VzA~and the 
  \ZNA~estimators for (a) $2<\pt<4~\gevc$ and (b) 
  $8<\pt<12~\gevc$. The average D-meson $\QpPb$ in $8<\pt<12~\gevc$ is 
  compared with the charged-particle $\QpPb$ calculated for $\pt>10~\gevc$~\cite{Adam:2014qja}. 
The vertical error bars and the empty boxes represent, respectively, the statistical and systematic uncertainties on the D-meson results.  
The filled boxes at $\QpPb=1$ indicate the correlated systematic uncertainties: the grey-filled box represents the uncertainty on the \pp~reference and the \pPb~analysis PID and track selection uncertainties, common to all estimators for a given $\pt$ interval; the red-filled box represents the correlated systematic uncertainty on $\Ncoll$ determination for the ZNA energy estimator. 
\label{fig:DQpPb3} 
} 
\end{center} 
\end{figure}

The average D-meson $\QpPb$ results obtained with the three estimators, for $2<\pt<4~\gevc$ and $8<\pt<12~\gevc$, are displayed as a function of centrality in Fig.~\ref{fig:DQpPb3}. The D-meson $\QpPb$ for $8<\pt<12~\gevc$ is compared with the analogous measurement for charged hadrons with $\pt>10~\gevc$~\cite{Adam:2014qja}. In this transverse momentum region also the production of charged hadrons is expected to scale with the number of binary nucleon--nucleon collisions~\cite{Adam:2014qja}. 
The measured trends of charged-particle $\QpPb$ at high $\pt$ in all the CL1 and V0A centrality classes were found to be reasonably described by an incoherent superposition of $\Ncoll$ \pp~collisions generated with PYTHIA, after defining the event centrality from the charged-particle multiplicity in the rapidity region covered by each estimator in the same way as in data ($|\eta|<1.4$ for CL1, $2.8<\eta<5.1$ for V0A)~\cite{Adam:2014qja}. 
 
The $\QpPb$ results for D mesons and charged hadrons with $\pt>10~\gevc$ show a similar trend as a function of centrality and estimator due to the bias in the centrality determination, as observed in~\cite{Adam:2014qja} based on high-$\pt$ particle production in the light flavour sector. The results presented in this paper allow these studies to be extended into the charm sector and down to low $\pt$.

\section{Relative yields as a function of multiplicity}
\label{sec:DvM}

$\Dzero$, $\Dplus$ and $\Dstar$ meson yields were also studied as a function of the charged-particle multiplicity in two pseudorapidity intervals, see Sec.~\ref{sec:DefRelativeMult}.  
The D-meson yields were evaluated for various multiplicity and $\pt$ intervals and the results are reported in terms of corrected per-event yields normalised to the multiplicity-integrated values 
\begin{equation} 
\frac{({\rm d}^2N^{\rm D}/{\rm d} y{\rm d}\pt)^j}{\langle{\rm d}^2N^{\rm D}/{\rm d} y{\rm d}\pt\rangle} 
= 
\left( \frac{1}{N_{\rm events}^j}\frac{N^j_{\rm raw \ D}}{\epsilon^j_{\rm prompt \ D}}\right) {\Bigg /}  
\left( \frac{1}{N_{\rm MB \ trigger}/\epsilon_{\rm MB \ trigger}} 
\frac{\langle N_{\rm raw \ D}\rangle}{\langle\epsilon_{\rm prompt \ D}\rangle} \right)\,, 
\label{eq:RelativeYieldsFormula} 
\end{equation} 
where the index $j$ identifies the multiplicity interval, $N_{\rm raw \ D}^j$ is the raw yield extracted from the fit to the invariant mass distribution in each multiplicity interval, $\epsilon_{\rm prompt \ D}^j$ represents the reconstruction and selection efficiencies for prompt D mesons, and $N^j_{\rm events}$ is the number of events analysed in each multiplicity interval. The efficiencies were estimated with Monte Carlo simulations (see Sec.~\ref{sec:Dreco}).  
Equation~(\ref{eq:RelativeYieldsFormula}) holds under the assumption that the relative contribution to the raw D-meson yield due to the feed-down from beauty-hadron decays does not depend on the multiplicity of the event, and therefore cancels out in the ratio to the multiplicity-integrated values.  
This assumption is justified by the beauty production measurements as a function of multiplicity in \pp~collisions, and also by PYTHIA simulations~\cite{Adam:2015ota}. 
The acceptance correction, defined as the fraction of D mesons within a given rapidity and $\pt$ interval that decay into pairs or triplets of particles within the detector coverage, cancels out in this ratio. The number of events used for the normalisation of the multiplicity-integrated yield must be corrected for the fraction of non-single diffractive events that are not accepted by the minimum-bias trigger condition, expressed as $N_{\rm MB \ trigger}/\epsilon_{\rm MB \ trigger}$ with $\epsilon_{\rm MB \ trigger}=(96.4 \pm 3.1)\%$~\cite{Abelev:2014epa}. It was verified with PYTHIA 6.4.21 Monte Carlo simulations that the minimum-bias trigger is 100\% efficient for D mesons in the kinematic range of the measurement, meaning that the number of D mesons in the minimum-bias triggered events is the same as in the sample of non-single diffractive events.

\subsection{Systematic uncertainties}
\label{sec:ntrksyst}

In this section the systematic uncertainties estimated for the D-meson measurements as a function of $\Ntrk$ and as a function of the $\Nvzero$ multiplicity are outlined. 
 
The most significant source of systematic uncertainty is the one related to the signal extraction procedure.  
The raw D-meson yields were obtained by fixing the position of the Gaussian signal peak to the world averages of the D-meson masses, and the widths to the values obtained from the fit to the multiplicity integrated invariant mass distributions.  
To estimate the yield extraction uncertainty the fit parameters were varied as described in Sec.~\ref{sec:Dreco}.  
In addition to the variations listed in Sec.~\ref{sec:Dreco}, the fits were performed also allowing the position and the width of the Gaussian terms to remain free in the individual multiplicity intervals. 
The yield extraction uncertainty was estimated based on the stability of the ratio of the raw yields $N_{\rm raw \ D}^j/\langle N_{\rm raw \ D}\rangle$, where the same raw yield extraction method was used in the multiplicity interval $j$ and for the multiplicity-integrated result.  
The magnitude of this uncertainty depends on $\pt$ and meson species. 
The contribution of the yield extraction procedure to the systematic uncertainties varied between 4--10\%. 
 
The influence of D-meson selections, due to the PID and the topological selections, were examined and found to have no significant effect on the final result, since they enter equally into the numerator and denominator of Eq.~(\ref{eq:RelativeYieldsFormula}).  
 
As mentioned in Sec.~\ref{sec:Dreco}, the contribution of feed-down from B decays to the raw yield was estimated based on FONLL calculations~\cite{Cacciari:2012ny}.  
In this case, it was assumed that the fraction of D mesons that are not from feed-down decays, \fprompt, remains constant as a function of multiplicity, causing it to cancel out in the numerator and denominator of the ratio in Eq.~(\ref{eq:RelativeYieldsFormula}).  
The feed-down contribution was therefore not explicitly subtracted from the final result.  
A systematic uncertainty related to this hypothesis was assigned by assuming that the fraction $f_{\rm B}^j/\langle f_{\rm B} \rangle$, where $f_{\rm B} = 1 - \fprompt$, increases linearly from $1/2$ to $2$ from the lowest to the highest multiplicity intervals. 
The resulting uncertainty depends on multiplicity, $\pt$ and meson species, and ranges from $^{+4}_{-0}$\% to $^{+10}_{-0}$\% at low multiplicity and from $^{+0}_{-4}$\% to $^{+0}_{-20}$\% at high multiplicity.  
 
In the analyses as a function of $\Ntrk$, the relative average values of $\Ntrk^j/\langle\Ntrk\rangle$ for each interval were corrected to give relative $(\dNdEta)^j/\langle\dNdEta\rangle$ values, as described in Sec.~\ref{sec:DefRelativeMult}.  
The systematic uncertainty due to this correction was estimated in the simulations based on the resolution and the linearity of the correlation between the number of tracklets, $\Ntrk$, and the number of generated charged primary particles, $\Nch$. 
The deviation from linearity was found to contribute by roughly 5\% to the uncertainty on the relative multiplicity.  
Finally, the uncertainty on the measured $\langle\dNdEta\rangle$ in inelastic \pPb~collisions measured in~\cite{ALICEpPbmult} was considered. This contributed an uncertainty of approximately 4\%. 
The total systematic uncertainty on the relative charged-particle density per $\Ntrk$ interval was found to be 6.3\%. 
 
In the analyses as a function of $\Nvzero$, the measurements are reported as a function of the relative multiplicity $\Nvzero \big/ \langle \Nvzero \rangle$. The uncertainty on the mean multiplicity values, $\Nvzero$, was determined by comparing the mean and median values of the distributions. It was found to be below 5\% for each multiplicity interval, and about $30\%$ for the multiplicity-integrated value.

\subsection{Results}
\label{sec:ntrkresults}

The relative D-meson yields were calculated for each $\pt$ and multiplicity interval according to Eq.~(\ref{eq:RelativeYieldsFormula}).  
The results are reported as a function of the relative charged-particle multiplicity at both backward and central rapidity.  
It is worth noting that the smaller number of reconstructed D mesons in the lowest and highest $\pt$ intervals\footnote{ 
The number of reconstructed D mesons in the lowest and highest $\pt$ intervals is smaller than in the other $\pt$ intervals. 
At low $\pt$, the strategy employed to cope with the low signal-to-background ratio was to apply tight topological selections, decreasing the selection efficiency and consequently the number of reconstructed D mesons.  
At high $\pt$, the small number of candidates is the consequence of the steeply falling D-meson $\pt$ spectra.  
} limited the number of multiplicity intervals of the measurement for those $\pt$ intervals.

\begin{figure}[!htb] 
\begin{center} 
\subfigure[D mesons with $2<\pt<4~\gev/c$]{ 
	\label{fig:DYieldCorr_2_4} 
	\includegraphics[width=0.45\columnwidth]{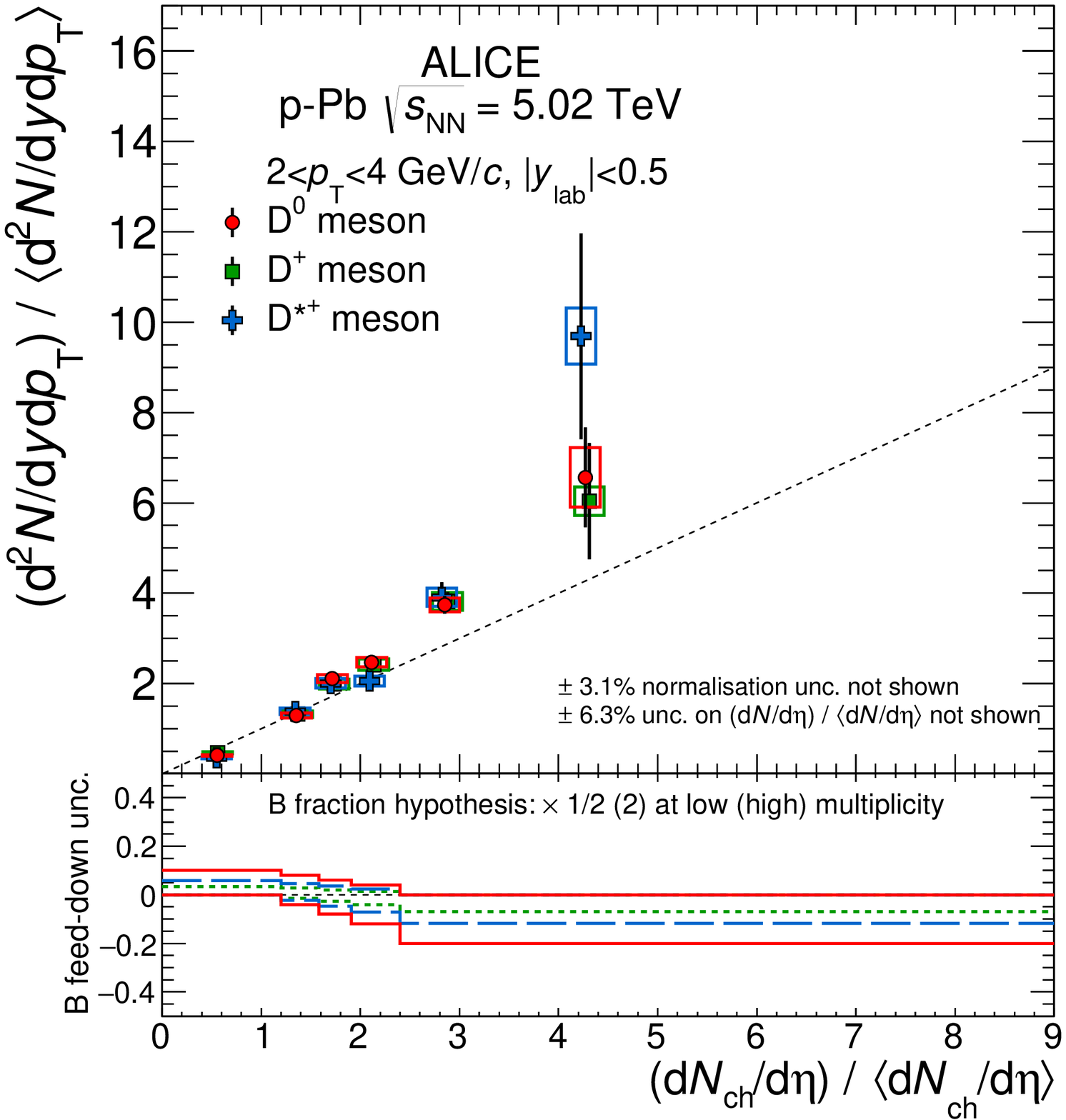} 
	}  
\subfigure[D mesons with $4<\pt<8~\gev/c$]{ 
	\label{fig:DYieldCorr_4_8} 
	\includegraphics[width=0.45\columnwidth]{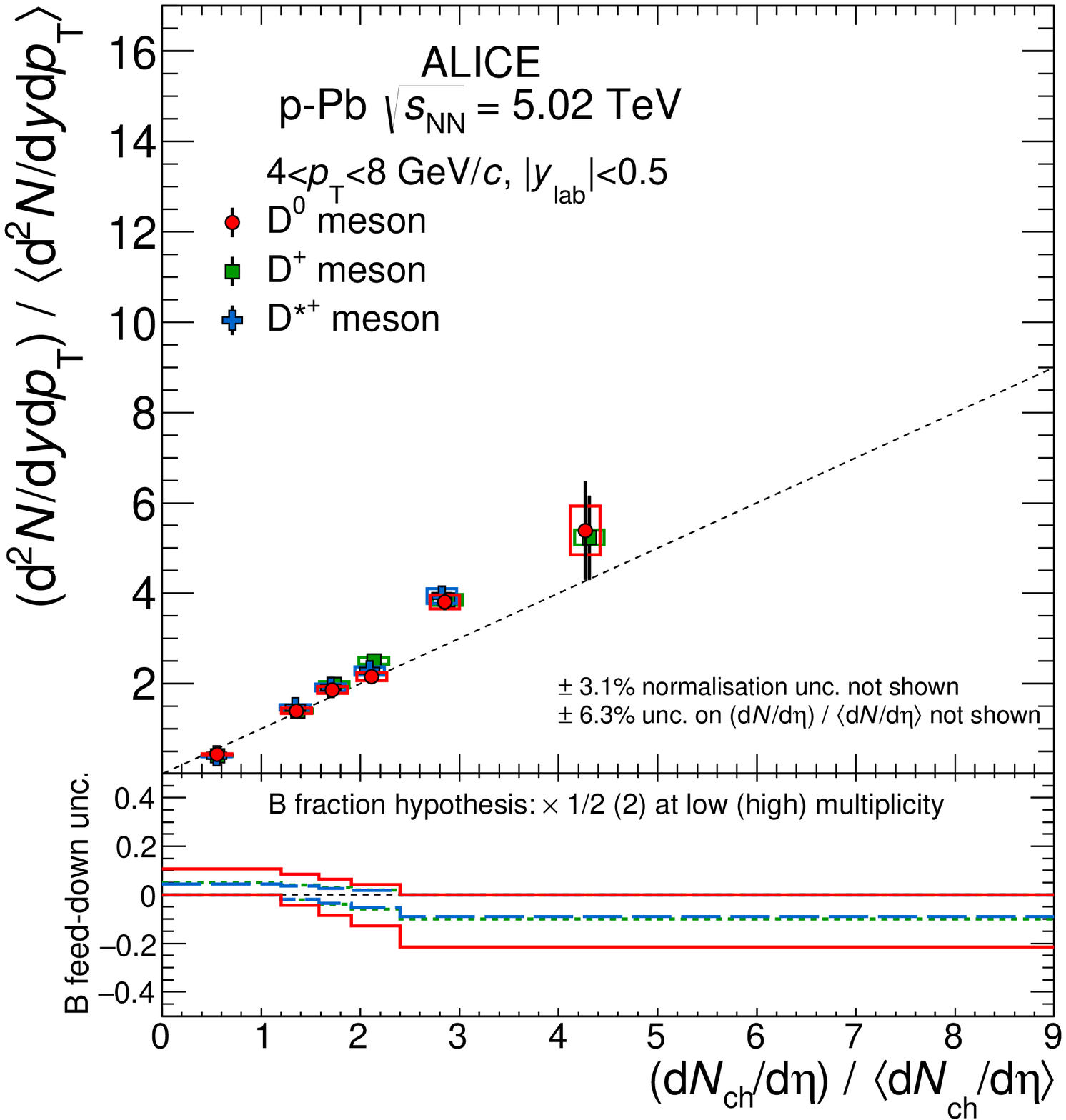} 
	}  
\caption{ 
Relative $\Dzero$, $\Dplus$ and $\Dstar$ meson yields for two selected $\pt$ intervals as a function of charged-particle multiplicity at central rapidity.  
The relative yields are presented in the top panels with their statistical (vertical bars) and systematic (empty boxes) uncertainties, apart from the feed-down fraction uncertainty, which is drawn separately in the bottom panels.  
The position of the points on the abscissa is the average value of $(\dNdEta) \big/ \langle \dNdEta \rangle$.  
For $\Dplus$ and $\Dstar$ mesons the points are shifted horizontally by $1.5\%$ to improve the visibility.  
The diagonal (dashed) line is also shown to guide the eye. 
\label{fig:DYieldCorr} 
} 
\end{center} 
\end{figure}

The relative $\Dzero$, $\Dplus$ and $\Dstar$ yields were measured in five $\pt$ intervals from $1$ to $24~\gevc$ as a function of the charged-particle multiplicity at mid-rapidity.  
Figure~\ref{fig:DYieldCorr} presents the measurements for selected $\pt$ intervals with their statistical (vertical bars) and systematic (boxes) uncertainties, apart from the feed-down fraction uncertainty, which is drawn separately in the bottom panels.  
The position of points on the abscissa is the average value of $(\dNdEta) \big/ \langle \dNdEta \rangle$, but for some meson species they are shifted horizontally by $1.5\%$ to improve the visibility.  
The relative yields of the three D-meson species are consistent with one another in all $\pt$ intervals within uncertainties. 
 
\begin{figure}[!htb] 
\begin{center} 
\subfigure[$\pt$ dependence]{ 
	\label{fig:DYieldCorrAv_pt} 
	\includegraphics[width=0.45\columnwidth]{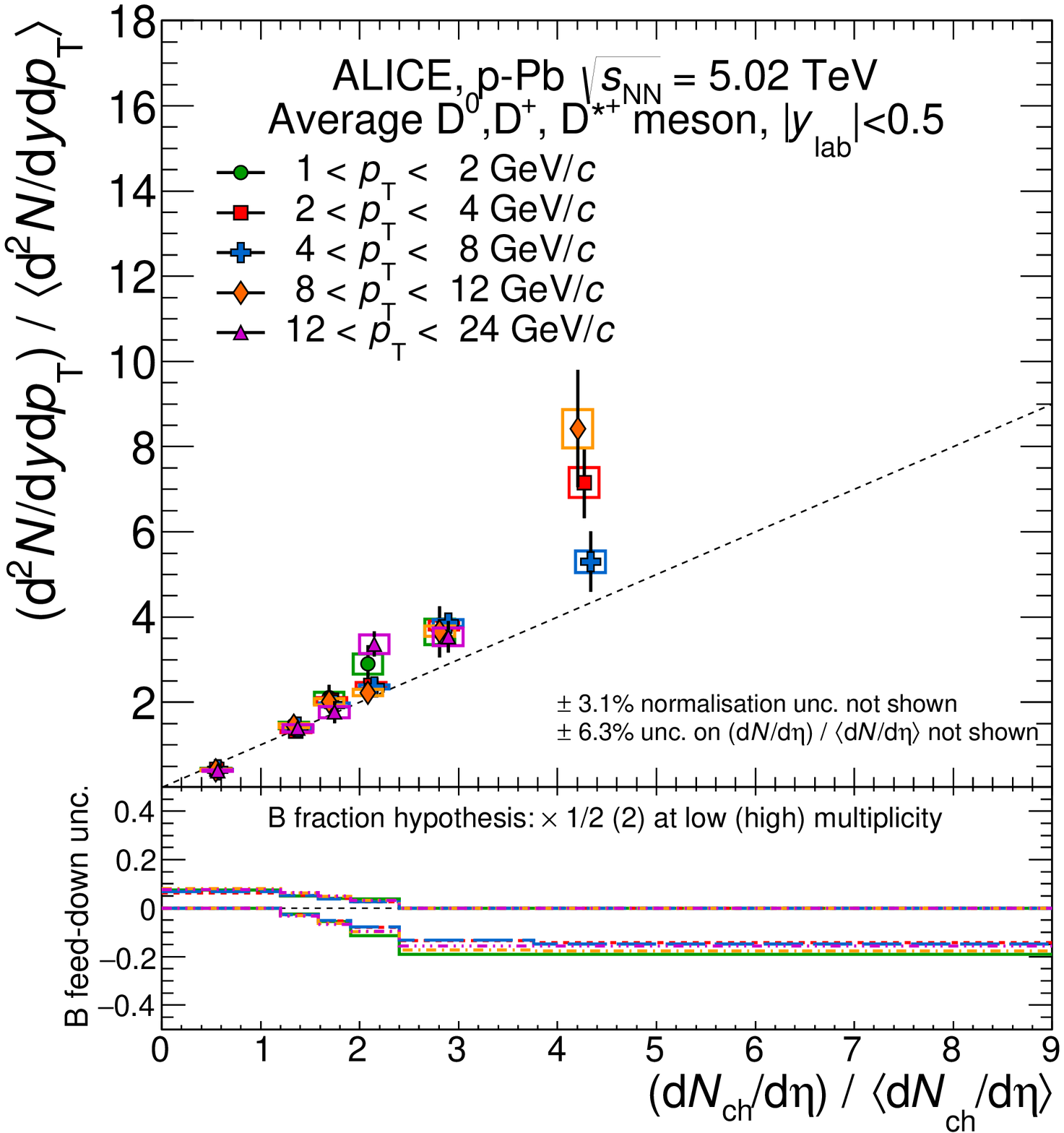} 
	}  
\subfigure[Ratios of $\pt$ intervals vs. $2<\pt<4~\gev/c$.]{ 
	\label{fig:DYieldCorrAv_pt_ratio} 
	\includegraphics[width=0.45\columnwidth]{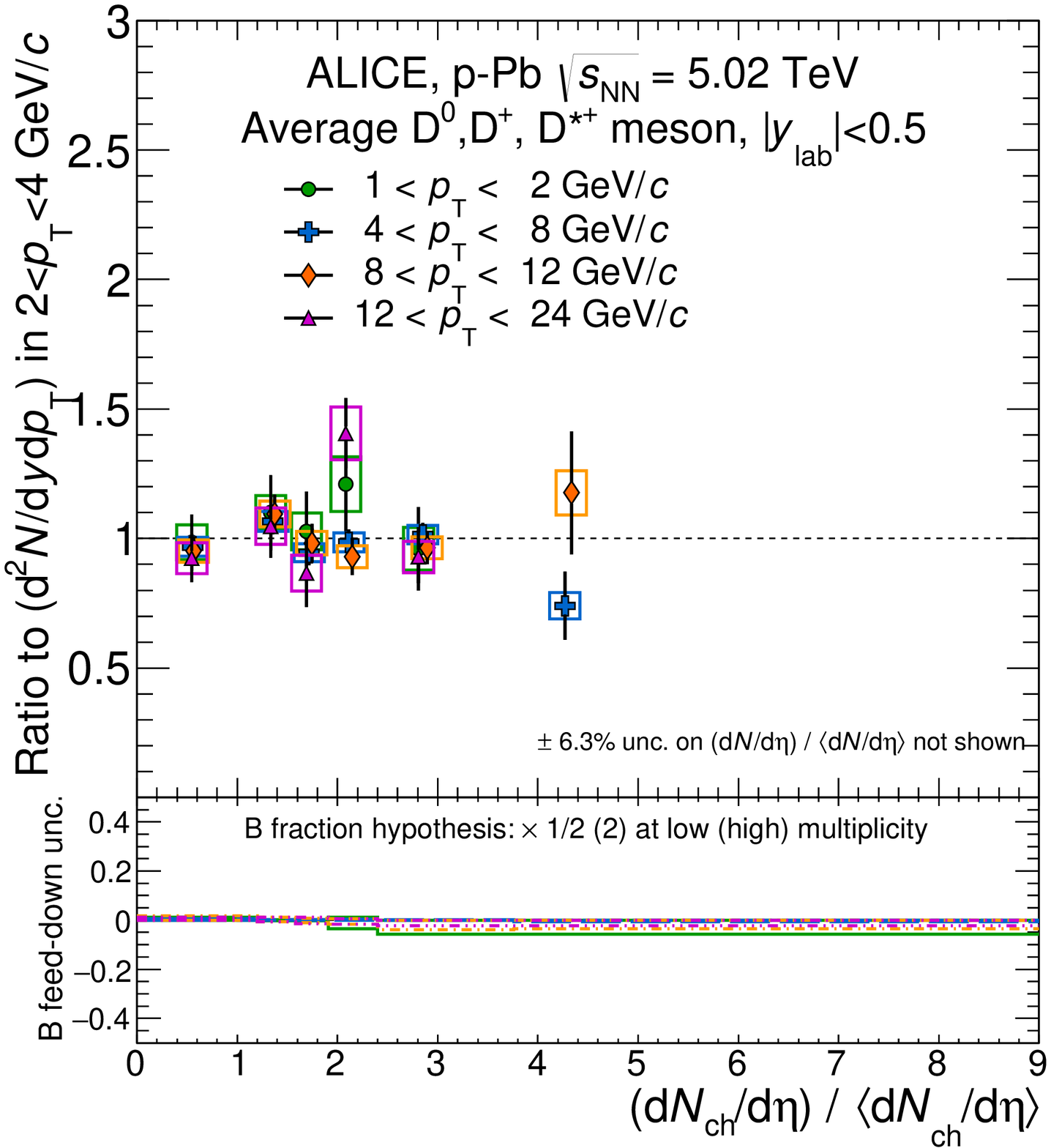} 
	}  
\caption{ 
Average of relative $\Dzero$, $\Dplus$ and $\Dstar$ yields as a function of the relative charged-particle multiplicity at central rapidity. 
	(a)~Average of relative D-meson yields in $\pt$ intervals.  
	(b)~Ratio of the average relative yields in all $\pt$ intervals with respect to that of the $2<\pt<4~\gev/c$ interval. 
	The results are presented in the top panels with their statistical (vertical bars) and systematic (boxes) uncertainties, apart from the feed-down fraction uncertainty, which is drawn separately in the bottom panels. 
	The position of the points on the abscissa is the average value of $(\dNdEta) \big/ \langle \dNdEta \rangle$.  
	For some $\pt$ intervals the points are shifted horizontally by $1.5\%$ to improve the visibility.  
	The dashed lines are also shown to guide the eye, a diagonal on (a) and a constant on (b). 
\label{fig:DYieldCorrAverage} 
} 
\end{center} 
\end{figure}

The average of the relative $\Dzero$, $\Dplus$ and $\Dstar$ yields was evaluated considering the inverse square of their relative statistical uncertainties as weights.  
The yield extraction uncertainties were treated as uncorrelated systematic uncertainties, while the feed-down subtraction uncertainties were considered as correlated uncertainty sources.  
Figure~\ref{fig:DYieldCorrAv_pt} presents the average D-meson yields for each $\pt$ interval.  
The results are reported in Table~\ref{tab:DAverageMult}.  
The $\pt$ evolution of the yields was examined using the results in the $2<\pt<4~\gevc$ interval as reference and by computing the ratio between the average relative D-meson yields in the various $\pt$ intervals and those in $2<\pt<4~\gevc$. The results are shown in Fig.~\ref{fig:DYieldCorrAv_pt_ratio}.  
The yield increase is independent of transverse momentum within the uncertainties of the measurement. 
The D-meson yields show a faster-than-linear increase with the charged-particle multiplicity at central rapidity.  
The yield increase is approximately a factor of 7 for multiplicities of 4.2 times $\langle \dNdEta \rangle$.  
These results are compared with the equivalent measurements in \pp~collisions, as well as with model calculations, in Sec.~\ref{sec:DvsMult:DataModel}. 

%
%
%
 
\begin{figure}[!htb] 
\begin{center} 
\subfigure[$\pt$ dependence]{ 
	\label{fig:DYieldCorrAv_Vza_pt} 
	\includegraphics[width=0.475\columnwidth]{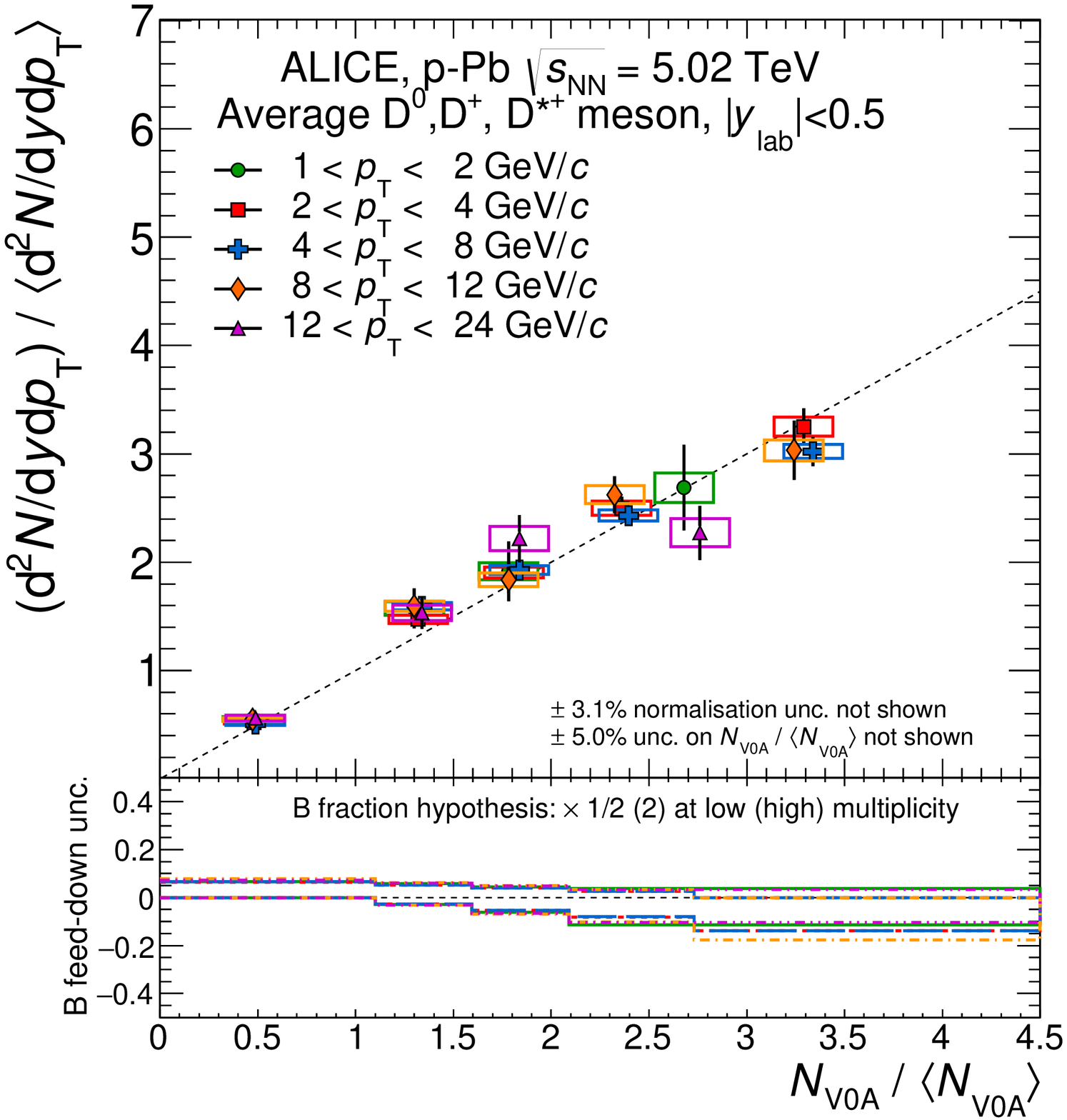} 
	}  
\subfigure[Ratios of $\pt$ intervals vs. $2<\pt<4~\gev/c$.]{ 
	\label{fig:DYieldCorrAv_Vza_pt_ratio} 
	\includegraphics[width=0.475\columnwidth]{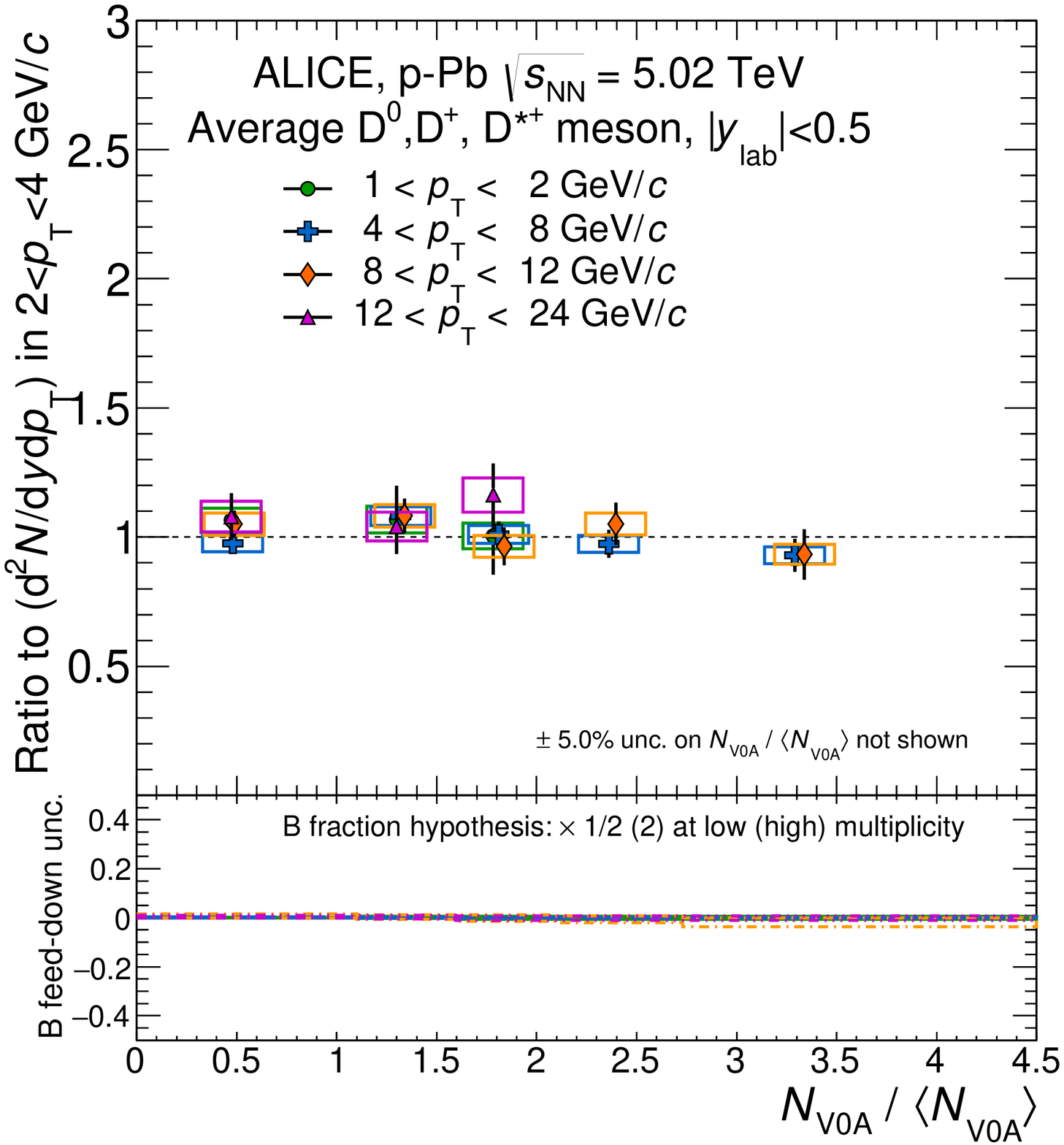} 
	}  
\caption{ 
Average of relative $\Dzero$, $\Dplus$ and $\Dstar$ yields as a function of the relative V0A multiplicity, $\Nvzero$, measured at $2.8 < \eta < 5.1$.  
The relative yields are presented in the top panels with their statistical (vertical bars) and systematic (boxes) uncertainties, apart from the uncertainty on the B feed-down fraction, which is drawn separately in the bottom panels. 
	The position of the points on the abscissa is the average value of $\Nvzero \big/ \langle \Nvzero \rangle$.  
	For some $\pt$ intervals the points are shifted horizontally by $1.5\%$ to improve the visibility.  
	The dashed lines are also shown to guide the eye, a diagonal on (a) and a constant on (b). 
\label{fig:DvsMultYieldCorrVzero} 
} 
\end{center} 
\end{figure}

The measurement of the relative $\Dzero$, $\Dplus$ and $\Dstar$ yields was also performed as a function of the relative charged-particle multiplicity at large rapidity in the Pb-going direction, thus introducing an $\eta$ gap between the regions  
where the D mesons and the multiplicity are measured. 
The charge collected by the V0A detector, $\Nvzero$, was considered as a multiplicity estimator (see Sec.~\ref{sec:DefRelativeMult}). Simulations have shown that the collected charge is proportional to the charged-particle multiplicity in the measured $\eta$ range, $2.8< \eta< 5.1$.  
The relative D-meson yields measured in $\pt$ and $\Nvzero$ intervals are reported as a function of the relative multiplicity in the V0A detector, $\Nvzero \big/ \langle \Nvzero \rangle$. 
The $\Dzero$, $\Dplus$ and $\Dstar$ yields are consistent with one another in all the measurement intervals, within uncertainties.  
The average D-meson yield was calculated with the same procedure used for the results as a function of charged-particle multiplicity at mid-rapidity.  
Figure~\ref{fig:DvsMultYieldCorrVzero} and Table~\ref{tab:DvsNvzero} summarise these measurements.  
The results are independent of transverse momentum within the uncertainties of the measurement. 
The charmed-meson yield increases with the multiplicity at backward rapidity.  
The yield increase is consistent with a linear growth as a function of multiplicity.  
The results as a function of V0A multiplicity indicate that the per-event D-meson yield increases as a function of multiplicity, regardless of the $\eta$ range in which the multiplicity is measured. This remains the case even when the charged-particle yield is measured in a different $\eta$ interval from the D mesons, which originate from the fragmentation of charm quarks produced in hard partonic scattering processes. 
 
One notable effect to consider when comparing the trends of D-meson production as a function of multiplicity at central and large rapidity is that the charged-particle multiplicity was observed to scale differently with the number of nucleons involved in the p--A interaction depending on $\eta$~\cite{Adare:2013nff,Adam:2014qja}. In particular, at central rapidity the charged-particle multiplicity is found to scale with the number of participant nucleons, $\Npart$, while at large rapidities in the Pb-going direction (i.e. in the V0A acceptance) it scales with the number of participants of the Pb nucleus, which is equal to $\Npart -1=\Ncoll$ in \pPb~collisions.  
 
It was verified that the results of the D-meson yields as a function of multiplicity are consistent with those of the $\QpPb$ analysis (see Sec.~\ref{sec:EA}).  
In the $\QpPb$ analysis, D-meson production is studied by dividing the events into centrality classes equally populated by 20\% of the events, whereas in this section we examine events with extremely high multiplicity (see Tables~\ref{tab:multrangesSPD} and~\ref{tab:multrangesvzero}).  
Events with low (high) multiplicity correspond to interactions with a smaller (larger) number of hard scatterings per nucleon-nucleon collision, as well as to negative (positive) multiplicity fluctuations which affect event classification and influence both measurements.

%
\subsubsection{Comparison of \pPb~data with \pp~results and models} 
\label{sec:DvsMult:DataModel}

\begin{figure}[!htb] 
\begin{center} 
\subfigure[Multiplicity at central rapidity]{ 
	\label{fig:DvsMultYield_pp_MidRap} 
	\includegraphics[width=0.475\columnwidth]{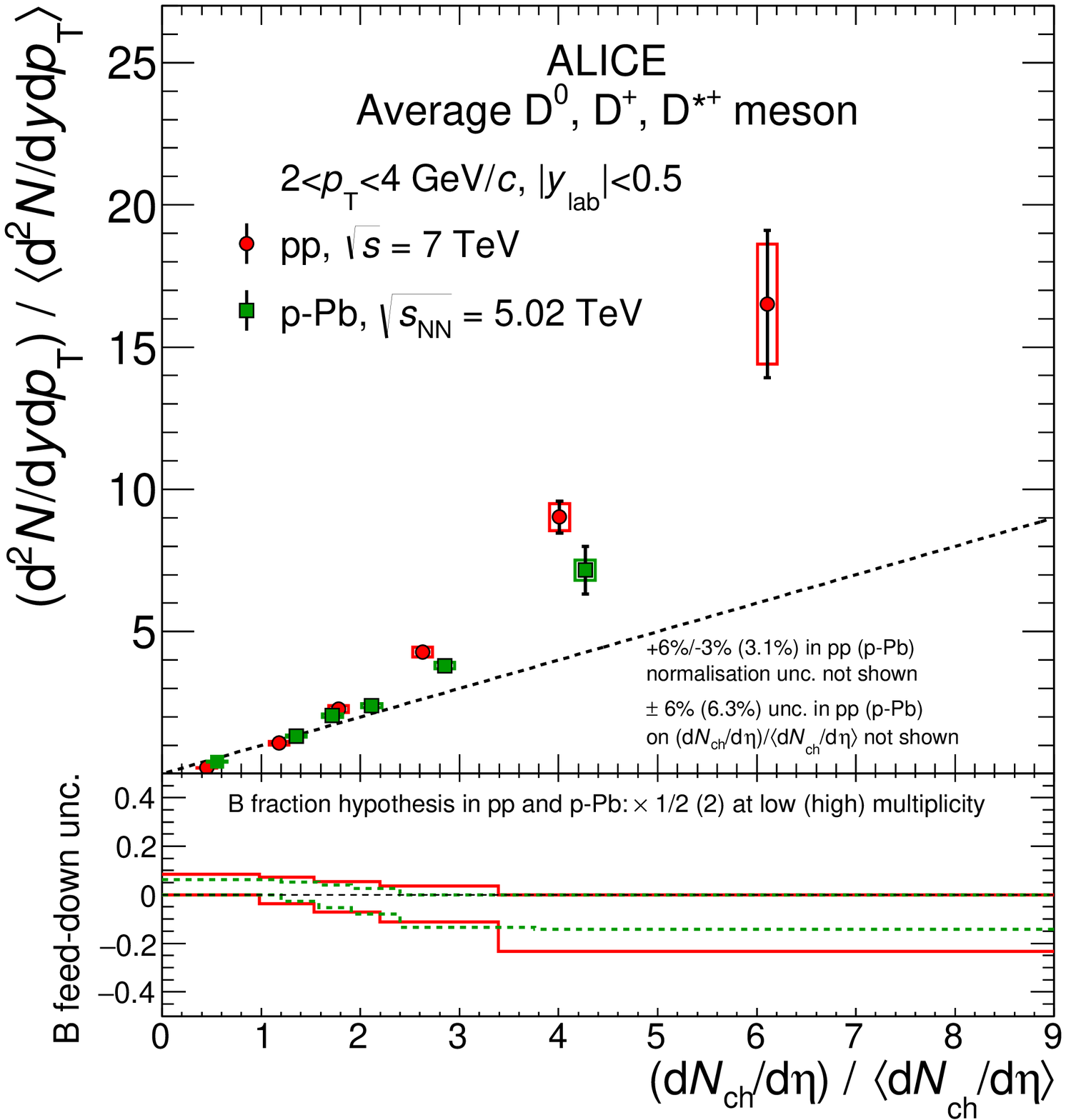} 
	}  
\subfigure[Multiplicity at backward rapidity]{ 
	\label{fig:DvsMultYield_pp_FwdRap} 
	\includegraphics[width=0.475\columnwidth]{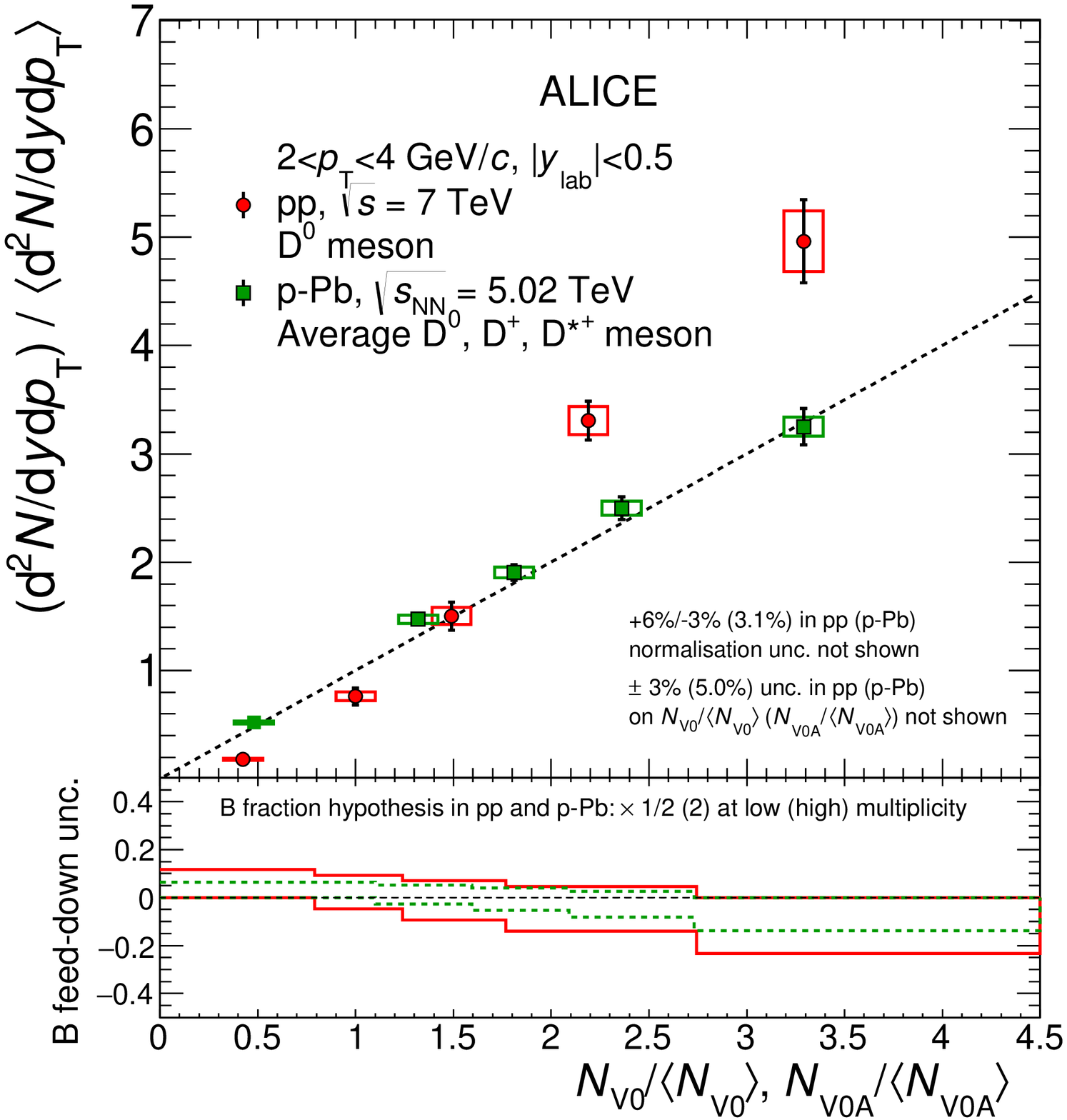} 
	}  
\caption{ 
Average relative D-meson yields in $|y_{\rm lab}|<0.5$ as a function of (a) the relative charged-particle multiplicity at mid-rapidity $|\eta|<1.0$, and (b) at backward-rapidity $2.8 < \eta < 5.1$ (including also $-3.7 < \eta < -1.7$ in \pp~data) for $2<\pt<4~\gevc$. 
The relative yields are presented in the top panels with their statistical (vertical bars) and systematic (boxes) uncertainties, apart from the uncertainty on the B feed-down fraction, which is drawn separately in the bottom panels. 
The positions of the points on the abscissa are the average values of $(\dNdEta) \big/ \langle \dNdEta \rangle$ or $\Nvzero \big/ \langle \Nvzero \rangle$. A diagonal (dashed) line is also shown to guide the eye. 
\label{fig:DvsMultYield_pp_pPb} 
} 
\end{center} 
\end{figure}

The relative D-meson yield (average of $\Dzero$, $\Dplus$ and $\Dstar$) as a function of charged-particle multiplicity at central rapidity in \pPb~collisions at $\sqrtsNN=5.02$~TeV is compared with the corresponding \pp~measurements at $\sqrts=7$~TeV for $2<\pt<4~\gevc$ in Fig.~\ref{fig:DvsMultYield_pp_MidRap}.  
A similar relative increase of charmed-meson yield with charged-particle multiplicity is observed in \pp~and \pPb~collisions.  
Note that the multiplicity is measured for both pp and \pPb~collisions in the same pseudorapidity range in the laboratory system, which corresponds to different ranges in the centre-of-mass frame for the two collision systems, due to the asymmetry of the beam energies in the \pPb~case. 
 
The increasing yield in \pp~data can be described by calculations taking into account the contribution of Multiple-Parton Interactions (MPI)~\cite{Bartalini:2010su,Sjostrand:1987su,Porteboeuf:2010dw},  
by the influence of the interactions between colour sources in the percolation model~\cite{Ferreiro:2012fb,Ferreiro:2015gea},  
or by the effect of the initial conditions of the collision followed by a hydrodynamic evolution computed with the EPOS~3 event generator~\cite{Drescher:2000ha,Werner:2013tya} where the individual scatterings are identified with parton ladders.  
In \pPb~collisions, the multiplicity dependence of heavy-flavour production is also affected by the presence of multiple binary nucleon--nucleon interactions, and the initial conditions of the collision are modified due to CNM effects. 
 
Charmed-meson yields in \pp~and \pPb~collisions as a function of the relative multiplicity at large rapidity are compared in Fig.~\ref{fig:DvsMultYield_pp_FwdRap} for $2<\pt<4~\gevc$.  
The multiplicity in \pPb~collisions is measured in $2.8 < \eta < 5.1$ in the Pb-going direction, whereas in \pp~data the multiplicities at backward ($2.8 < \eta < 5.1$) and forward ($-3.7 < \eta < -1.7$) pseudorapidity were summed together.  
The D-meson yields increase faster in \pp~than \pPb~collisions as a function of the relative multiplicity at backward rapidity.  
The different pseudorapidity intervals of the multiplicity measurement may contribute to this observation. %
In addition, measurements in \pPb~collisions differ from those in \pp~interactions because the initial conditions of the collision are affected by the presence of the Pb nucleus, and because there are multiple binary nucleon--nucleon interactions per \pPb~collision.  

\begin{figure}[!htb] 
\begin{center} 
\includegraphics[width=0.75\columnwidth]{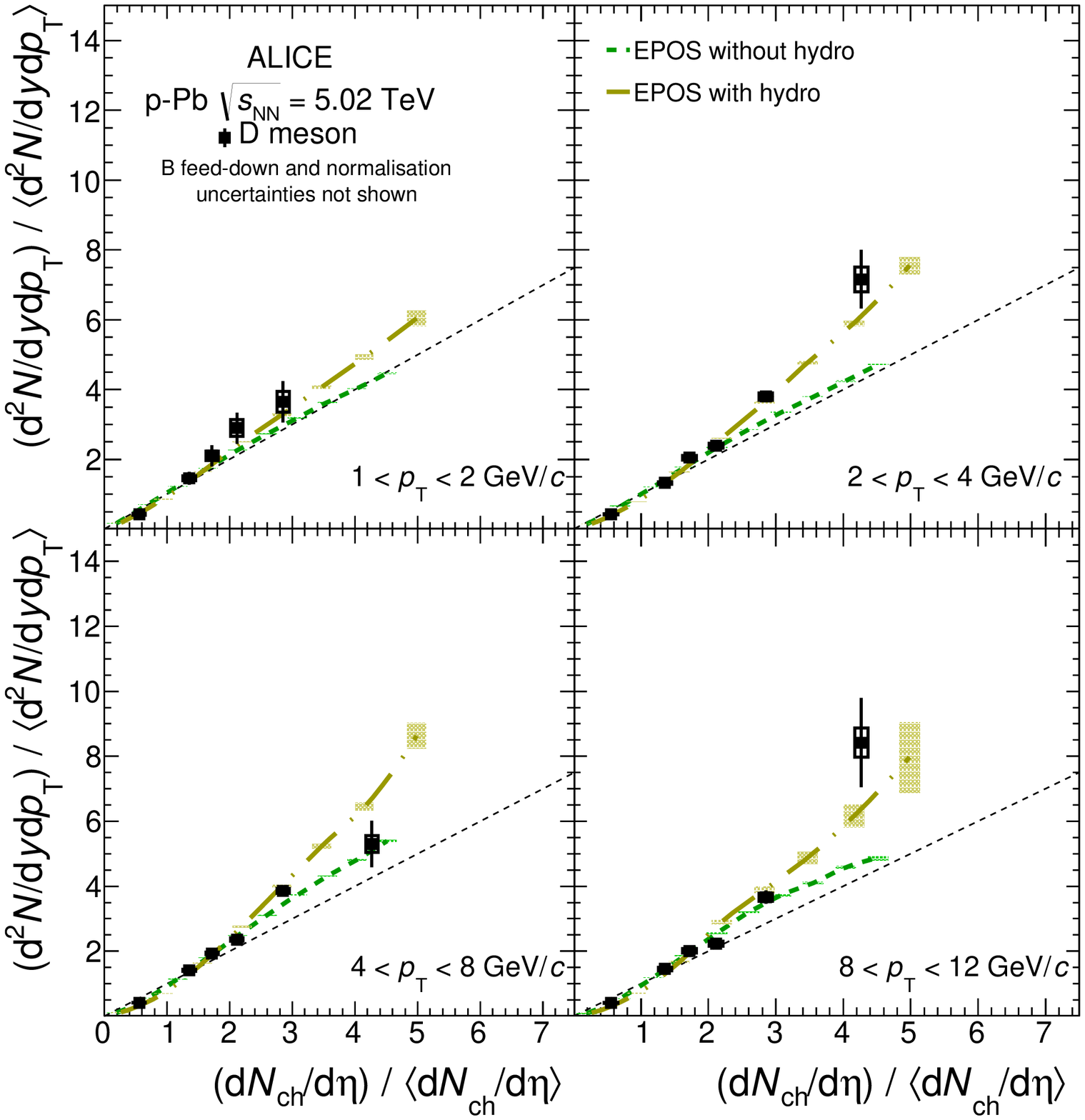} 
\caption{ 
\label{fig:DvsMult_theory_summary} 
Average relative D-meson yield as a function of the relative charged-particle multiplicity at central rapidity in different $\pt$ intervals.  
The systematic uncertainties on the data normalisation ($\pm3.1\%$), on the $(\dNdEta) \Big/ \langle \dNdEta \rangle$ values ($\pm6.3\%$), and on the feed-down contribution are not shown in this figure. 
The calculations of EPOS~3.116 with and without hydro~\cite{Drescher:2000ha,Werner:2013tya} are also shown.  
The coloured lines represent the calculation curves, whereas the shaded bands represent their statistical uncertainties at given values of $(\dNdEta) \Big/ \langle \dNdEta \rangle$.  
A diagonal (dashed) line is also shown to guide the eye. 
} 
\end{center} 
\end{figure}

\begin{figure}[!htb] 
\begin{center} 
\includegraphics[width=0.75\columnwidth]{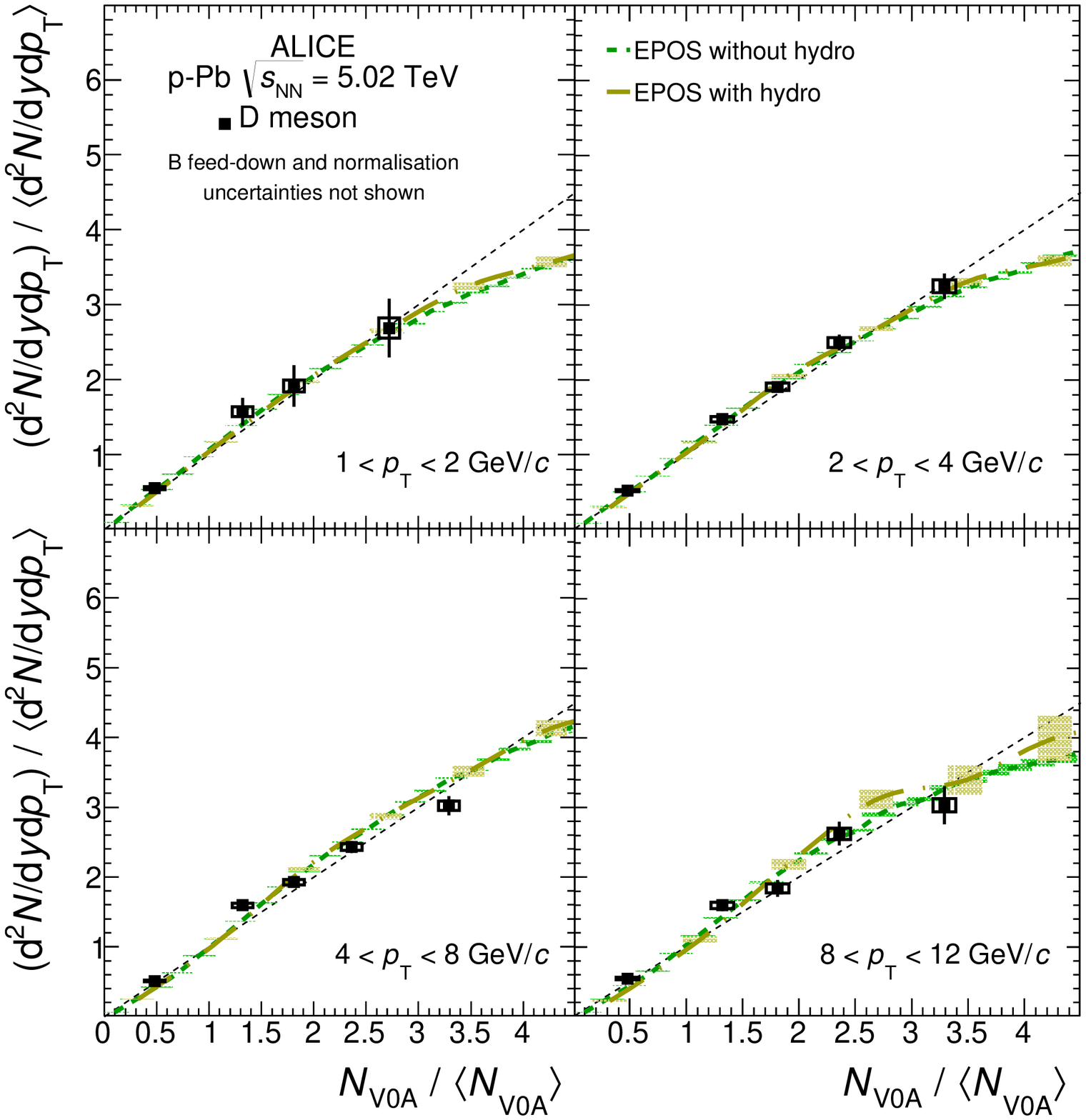} 
\caption{ 
\label{fig:DvsMult_theory_V0A_summary} 
Average relative D-meson yield as a function of the relative V0A multiplicity at backward rapidity in different $\pt$ intervals.  
The systematic uncertainties on the data normalisation ($\pm3.1\%$), on the $\Nvzero \big/ \langle \Nvzero \rangle$ values ($\pm5.0\%$), and on the feed-down contribution are not shown in this figure. 
The calculations of EPOS~3.116 with and without hydro~\cite{Drescher:2000ha,Werner:2013tya} are also shown.  
The coloured lines represent the calculation curves, whereas the shaded bands represent their statistical uncertainties at given values of $N_\mathrm{V0A} \Big/ \langle N_\mathrm{V0A} \rangle$.  
A diagonal (dashed) line is also shown to guide the eye. 
} 
\end{center} 
\end{figure}

Figures~\ref{fig:DvsMult_theory_summary} and~\ref{fig:DvsMult_theory_V0A_summary} present comparisons of the D-meson results and EPOS~3.116 model estimates.  
The EPOS~3 event generator~\cite{Drescher:2000ha,Werner:2013tya} imposes the same theoretical framework for various colliding systems: \pp, p--A and A--A.  
The initial conditions are generated using the ``Parton-based Gribov-Regge" formalism~\cite{Drescher:2000ha} of multiple scatterings. Each individual scattering is identified with a parton ladder, composed of a pQCD hard process with initial- and final-state radiation.  
The non-linear effects of parton evolution are treated introducing a saturation scale below which those effects become important.  
With these initial conditions, a 3D+1 viscous hydrodynamical evolution is applied to the core of the collision~\cite{Werner:2013tya}.  
The measurements agree with the EPOS~3 model calculations within uncertainties. The results at high multiplicity are better reproduced by the calculation including a viscous hydrodynamical evolution of the collision, which predicts a faster-than-linear increase of the charmed-meson yield with multiplicity at central rapidity.  
The same calculation evaluates an approximately linear increase of the charmed-meson yield with the multiplicity measured at backward rapidity due to the reduced influence of flow on charged particles produced at large rapidity.  

\section{Summary}
\label{sec:conclusions}

The production of $\Dzero$, $\Dplus$ and $\Dstar$ mesons as a function of multiplicity in \pPb~collisions at $\sqrtsNN=5.02$~TeV, measured with the ALICE detector, has been reported.  
D mesons were reconstructed in their hadronic decays in different transverse momentum intervals within $1<\pt<24~\gevc$, in the centre-of-mass rapidity range $-0.96< y_{\mathrm{cms}}<0.04$.  
The multiplicity dependence of D-meson production was studied both by comparing their yields in \pPb~collisions for various centrality classes with those of binary scaled \pp~collisions at the same centre-of-mass energy via the nuclear modification factor, and by evaluating the relative yields sliced in multiplicity intervals with respect to the multiplicity-integrated ones.

The $\pt$-differential nuclear modification factor, $\QpPb$, of the D mesons was evaluated with three centrality estimators according to the multiplicity measured in different pseudorapidity intervals: \CL~in $|\eta|<1.4$, \VzA~in $2.8<\eta<5.1$ in the Pb-going direction, and the energy of slow neutrons detected by the \ZNA~calorimeter at very large rapidity.  
For each estimator, the events were classified in four classes corresponding to percentiles of the cross section: 0--20\%, 20--40\%, 40--60\% and 60--100\%.  
The $\QpPb$ results for the three D-meson species fluctuate around unity and are consistent in the measured $\pt$ and centrality intervals within uncertainties.  
The results with the \CL~estimator suggest an ordering from higher ($>1$) to lower ($<1$) $\QpPb$ values from the 0--20\% to the 60--100\% centrality class.  
This disparity is reduced when $\QpPb$ is calculated using the \VzA~estimator, and vanishes when it is determined with the \ZNA~estimator ($\QpPb \approx 1$).  
These effects are understood to be due to the biases in the centrality determination in \pPb~collisions based on measurements of multiplicity. 
The \ZNA~estimator is the least affected by these sources of biases, and the 
$\QpPb$ results obtained with this estimator indicate that there is no evidence of a centrality dependence of the D-meson production in \pPb~collisions with respect to that of \pp~collisions at the same centre-of-mass energy in the measured $\pt$ interval within the uncertainties.

The D-meson yields were also studied in \pPb~collisions as a function of the relative charged-particle multiplicity at mid-rapidity, $|\eta|<1.0$, and at large rapidity, $2.8<\eta<5.1$, in the Pb-going direction. 
The relative yields, i.e.\ the yields in a given multiplicity interval divided by the multiplicity-integrated ones, were calculated differentially in transverse momentum.  
In contrast to $\QpPb$, which examines particle production in samples of 20\% of the analysed events, this observable explores events from low to extremely high multiplicities corresponding to only 5\% (1\%) of the analysed events in \pPb~(\pp) collisions.  
The measurements of the relative yields for $\Dzero$, $\Dplus$ and $\Dstar$ mesons are consistent within the uncertainties. 
The D-meson yields increase with charged-particle multiplicity, and the increase is independent of $\pt$ within the measurement uncertainties.  
The yield increases with a faster-than-linear trend as a function of the charged-particle multiplicity at mid-rapidity.  
This behaviour is similar to that of the corresponding measurements in \pp~collisions at $\sqrts=7$~TeV.  
Possible interpretations include short-distance gluon radiation, contributions from Multiple-Parton Interactions, the influence of initial conditions followed by a hydrodynamic expansion (EPOS~3 event generator), or the percolation model scenario.  
In addition, the contribution from multiple binary nucleon-nucleon collisions must be considered in \pPb~collisions.  
By contrast, the increase of the charmed-meson yields as a function of charged-particle multiplicity at large rapidity in the Pb-going direction is consistent with a linear growth as a function of multiplicity.  
EPOS~3 Monte Carlo calculations are in reasonable agreement with the \pPb~results within uncertainties.  

\newenvironment{acknowledgement}{\relax}{\relax}
\begin{acknowledgement}
\section*{Acknowledgements}
The ALICE Collaboration would like to thank K. Werner, co-author of EPOS~3, for fruitful discussions and for providing their theoretical calculations. 
\input{acknowledgements.tex}    
\end{acknowledgement}

\bibliographystyle{utphys}
\bibliography{DvsMultpPb.bib}

\newpage
\appendix

\section{Result tables}

\begin{table}[!htp] 
{\footnotesize 
\begin{center} 
\begin{tabular}{lcccc} 
\hline 
 &  \multicolumn{4}{c}{ZNA estimator} \\ 
& 0--20\% & 20--40\% & 40--60\% & 60--100\%\\[1.01ex] \hline 
$\pt$~($\gevc$)&  \multicolumn{4}{c}{$\QpPb$} \\[1.01ex] 
1--2 & 
	$ 0.73 \pm 0.14 _{-0.16}^{+0.17}$ & 
	$ 0.82\pm 0.15 _{-0.19}^{+0.22}$ & 
	$ 0.63\pm 0.14 _{-0.14}^{+0.16}$ & 
	$ 0.85\pm 0.19 _{-0.22}^{+0.20}$  
	 \\[1.01ex]%
2--4 &  
	$ 1.02\pm 0.07 _{-0.19}^{+0.17}$ & 
	$ 1.03\pm 0.07 _{-0.19}^{+0.17}$ & 
	$ 0.93\pm 0.07 _{-0.17}^{+0.16}$ & 
	$ 0.87\pm 0.06 _{-0.16}^{+0.14}$  
	\\[1.01ex]%
4--6 &  
	$ 1.07\pm 0.06 _{-0.17}^{+0.16}$ & 
	$ 1.05\pm 0.06 _{-0.17}^{+0.16}$ & 
	$ 0.93\pm 0.06 _{-0.15}^{+0.15}$ & 
	$ 0.92\pm 0.06 _{-0.15}^{+0.14}$  
	 \\[1.01ex]%
6--8 & 	 
	$ 1.04\pm 0.08 _{-0.17}^{+0.16}$ & 
	$ 0.97\pm 0.08 _{-0.15}^{+0.15}$ & 
	$ 0.99\pm 0.08 _{-0.16}^{+0.15}$ & 
	$ 1.01\pm 0.09 _{-0.16}^{+0.15}$  
	 \\[1.01ex]%
8--12 &  
	$ 0.98\pm 0.08 _{-0.16}^{+0.16}$ & 
	$ 0.98\pm 0.08 _{-0.16}^{+0.16}$ & 
	$ 0.84\pm 0.09 _{-0.14}^{+0.14}$ & 
	$ 0.94\pm 0.09 _{-0.16}^{+0.15}$  
	 \\[1.01ex]%
12--16 & 
	$ 1.02\pm 0.19  _{-0.18}^{+0.17}$ & 
	$ 1.14\pm 0.23  _{-0.20}^{+0.19}$ & 
	$ 0.75\pm 0.17 _{-0.14}^{+0.14}$ & 
	$ 0.88\pm 0.19 _{-0.16}^{+0.16}$  
	 \\[1.01ex]%
16--24 & 
	$ 0.84\pm 0.37  _{-0.38}^{+0.28}$ & 
	$ 0.95\pm 0.25  _{-0.25}^{+0.22}$ & 
	$ 1.15\pm 0.41 _{-0.52}^{+0.39}$ & 
	--  
	 \\[1.01ex]%
\hline 
Normalisation unc. & $\pm 0.07$ & $\pm 0.05$ & $\pm 0.07$ & $\pm 0.08$ 
	 \\[1.01ex]%
\hline 
\end{tabular} 
\caption{Average \QpPb~of $\Dzero$, $\Dplus$ and $\Dstar$ mesons for the sum of particles and antiparticles in several multiplicity and $\pt$ intervals for \pPb~collisions at~$\sqrtsNN=5.02$~TeV as a function of the multiplicity at central rapidity evaluated with the ZNA estimator.  
The values are reported together with their uncertainties,  which are quoted as statistical followed by systematic uncertainties.  
\label{tab:DQpPbZNA} 
} 
\end{center} 
 
} 
\end{table}

\begin{table}[!htp] 
{\footnotesize 
\begin{center} 
\begin{tabular}{lcccc} 
\hline 
 &  \multicolumn{4}{c}{\CL~estimator} \\ 
& 0--20\% & 20--40\% & 40--60\% & 60--100\%\\[1.01ex] \hline 
$\pt$~($\gevc$)&  \multicolumn{4}{c}{$\QpPb$} \\[1.01ex] 
1--2 & 
	$ 0.90 \pm 0.16 _{-0.20}^{+0.22}$ & 
	$ 0.83 \pm 0.15 _{-0.23}^{+0.24}$ & 
	$ 0.50\pm 0.10 _{-0.13}^{+0.14}$ & 
	$ 0.53\pm 0.11 _{-0.15}^{+0.15}$  
	 \\[1.01ex]%
2--4 &  
	$ 1.35\pm 0.08 _{-0.24}^{+0.22}$ & 
	$ 0.89\pm 0.06 _{-0.16}^{+0.15}$ & 
	$ 0.78\pm 0.05 _{-0.15}^{+0.13}$ & 
	$ 0.50\pm 0.04 _{-0.09}^{+0.09}$  
	\\[1.01ex]%
4--6 &  
	$ 1.38 \pm 0.07 _{-0.22}^{+0.21}$ & 
	$ 0.99\pm 0.05 _{-0.16}^{+0.15}$ & 
	$ 0.73\pm 0.04 _{-0.12}^{+0.11}$ & 
	$ 0.47\pm 0.03 _{-0.08}^{+0.07}$  
	 \\[1.01ex]%
6--8 & 	 
	$ 1.28 \pm 0.09 _{-0.20}^{+0.20}$ & 
	$ 0.98\pm 0.07 _{-0.16}^{+0.15}$ & 
	$ 0.77\pm 0.06 _{-0.12}^{+0.12}$ & 
	$ 0.51\pm 0.05 _{-0.08}^{+0.08}$  
	 \\[1.01ex]%
8--12 &  
	$ 1.28 \pm 0.10 _{-0.21}^{+0.20}$ & 
	$ 0.96\pm 0.08 _{-0.16}^{+0.15}$ & 
	$ 0.69\pm 0.07 _{-0.11}^{+0.11}$ & 
	$ 0.47\pm 0.07 _{-0.08}^{+0.08}$  
	 \\[1.01ex]%
12--16 & 
	$ 1.19 \pm 0.22  _{-0.21}^{+0.20}$ & 
	$ 1.14\pm 0.21  _{-0.20}^{+0.20}$ & 
	$ 0.78\pm 0.16 _{-0.14}^{+0.14}$ & 
	--  
	 \\[1.01ex]%
16--24 & 
	$ 1.20 \pm 0.26 _{-0.34}^{+0.27}$ & 
	$ 1.24\pm 0.56  _{-0.28}^{+0.28}$ & 
	-- & 
	-- 
	 \\[1.01ex]%
\hline 
Normalisation unc. & $\pm 0.05$ & $\pm 0.05$ & $\pm 0.07$ & $\pm 0.23$ 
	 \\[1.01ex]%
\hline 
\end{tabular} 
\caption{Average \QpPb~of $\Dzero$, $\Dplus$ and $\Dstar$ mesons for the sum of particles and antiparticles in several multiplicity and $\pt$ intervals for \pPb~collisions at~$\sqrtsNN=5.02$~TeV as a function of the multiplicity at central rapidity evaluated with the CL1 estimator.  
The values are reported together with their uncertainties, which are quoted as statistical followed by systematic uncertainties.  
\label{tab:DQpPbCL1} 
} 
\end{center} 
 
} 
\end{table}

\begin{table}[!htp] 
{\footnotesize 
\begin{center} 
\begin{tabular}{lcccc} 
\hline 
 &  \multicolumn{4}{c}{V0A estimator} \\ 
& 0--20\% & 20--40\% & 40--60\% & 60--100\%\\[1.01ex] \hline 
$\pt$~($\gevc$)&  \multicolumn{4}{c}{$\QpPb$} \\[1.01ex] 
1--2 & 
	$ 0.86\pm 0.16 _{-0.19}^{+0.20}$ & 
	$ 0.70\pm 0.14 _{-0.19}^{+0.20}$ & 
	$ 0.77\pm 0.15 _{-0.23}^{+0.24}$ & 
	$ 0.65\pm 0.15 _{-0.19}^{+0.19}$  
	 \\[1.01ex]%
2--4 &  
	$ 1.21 \pm 0.08 _{-0.22}^{+0.20}$ & 
	$ 0.94\pm 0.06 _{-0.17}^{+0.15}$ & 
	$ 0.89\pm 0.06 _{-0.16}^{+0.15}$ & 
	$ 0.71\pm 0.06 _{-0.14}^{+0.13}$  
	\\[1.01ex]%
4--6 &  
	$ 1.19\pm 0.06 _{-0.19}^{+0.18}$ & 
	$ 1.06\pm 0.06 _{-0.17}^{+0.16}$ & 
	$ 0.89\pm 0.05 _{-0.14}^{+0.13}$ & 
	$ 0.75\pm 0.05 _{-0.12}^{+0.12}$  
	 \\[1.01ex]%
6--8 & 	 
	$ 1.08\pm 0.08 _{-0.17}^{+0.16}$ & 
	$ 1.04\pm 0.08 _{-0.16}^{+0.16}$ & 
	$ 0.99\pm 0.08 _{-0.16}^{+0.15}$ & 
	$ 0.76\pm 0.07 _{-0.12}^{+0.12}$  
	 \\[1.01ex]%
8--12 &  
	$ 1.14\pm 0.09 _{-0.19}^{+0.18}$ & 
	$ 0.92\pm 0.08 _{-0.15}^{+0.15}$ & 
	$ 0.92\pm 0.09 _{-0.15}^{+0.15}$ & 
	$ 0.75\pm 0.08 _{-0.12}^{+0.12}$  
	 \\[1.01ex]%
12--16 & 
	$ 1.04\pm 0.21  _{-0.18}^{+0.18}$ & 
	$ 1.09\pm 0.21  _{-0.19}^{+0.19}$ & 
	$ 0.99\pm 0.20 _{-0.19}^{+0.19}$ & 
	$ 0.81\pm 0.21 _{-0.15}^{+0.14}$  
	 \\[1.01ex]%
16--24 & 
	$ 1.06\pm  0.23  _{-0.35}^{+0.27}$ & 
	$ 1.08\pm  0.49 _{-0.26}^{+0.26}$ & 
	-- & 
	--  
	 \\[1.01ex]%
\hline 
Normalisation unc. & $\pm 0.05$ & $\pm 0.05$ & $\pm 0.06$ & $\pm 0.22$ 
	 \\[1.01ex]%
\hline 
\end{tabular} 
\caption{Average \QpPb~of $\Dzero$, $\Dplus$ and $\Dstar$ mesons for the sum of particles and antiparticles in several multiplicity and $\pt$ intervals for \pPb~collisions at~$\sqrtsNN=5.02$~TeV as a function of the multiplicity at central rapidity evaluated with the V0A estimator.  
The values are reported together with their uncertainties, which are quoted as statistical followed by systematic uncertainties.  
\label{tab:DQpPbV0A} 
} 
\end{center} 
 
} 
\end{table}

\begin{landscape} 
 
\begin{table}[!htp] 
{\footnotesize 
\begin{center} 
 
\begin{tabular}{lcccccc} 
\hline 
 	&  \multicolumn{6}{c}{$(\dNdEta) \big/ \langle \dNdEta \rangle$} \\[1.01ex] &  
	$ 0.56 \pm 0.04 $ & 
	$ 1.35 \pm 0.09$ & 
	$ 1.72 \pm 0.11  $ & 
	$ 2.11 \pm 0.13$ & 
	$ 2.85 \pm 0.18$ & 
	$ 4.27 \pm 0.27$ \\[1.01ex] 
\hline 
$\pt$~($\gevc$)&  \multicolumn{6}{c}{${\rm d}^2N/{\rm d}y{\rm d}p_{\rm T} \big/ \langle {\rm d}^2N/{\rm d}y{\rm d}p_{\rm T} \rangle $} \\[1.01ex] 
1--2 &  
	$ 0.42 \pm 0.04  \pm 0.03 _{-0.00}^{+0.03}$ & 
	$ 1.46 \pm 0.18  \pm 0.08 _{-0.04}^{+0.07}$ & 
	$ 2.10 \pm 0.30  \pm 0.13 _{-0.11}^{+0.09}$ & 
	$ 2.90 \pm 0.44  \pm 0.24 _{-0.33}^{+0.11}$ & 
	$ 3.65 \pm 0.60  \pm 0.30 _{-0.70}^{+0.00}$ & 
	 --\\[1.01ex]%
2--4 &  
	$ 0.43 \pm 0.01 \pm 0.01  _{-0.00}^{+0.03}$ & 
	$ 1.33 \pm 0.05 \pm 0.04 _{-0.03}^{+0.07}$ & 
	$ 2.05 \pm 0.08 \pm 0.06 _{-0.11}^{+0.08}$ & 
	$ 2.40 \pm 0.09 \pm 0.07 _{-0.19}^{+0.06}$ & 
	$ 3.80 \pm 0.13 \pm 0.11 _{-0.51}^{+0.00}$ & 
	$ 7.16 \pm 0.84 \pm 0.35 _{-1.02}^{+0.00}$  
	 \\[1.01ex]%
4--8 &  
	$ 0.42 \pm 0.01 \pm 0.01 _{-0.00}^{+0.03}$ & 
	$ 1.41 \pm 0.04 \pm 0.03 _{-0.04}^{+0.07}$ & 
	$ 1.93 \pm 0.06 \pm 0.04 _{-0.10}^{+0.08}$ & 
	$ 2.36 \pm 0.08 \pm 0.05 _{-0.18}^{+0.06}$ & 
	$ 3.86 \pm 0.11 \pm 0.08 _{-0.51}^{+0.00}$ & 
	$ 5.30 \pm 0.71 \pm 0.25 _{-0.79}^{+0.00}$  
	 \\[1.01ex]%
8--12 &  
	$ 0.41 \pm 0.02 \pm 0.01 _{-0.00}^{+0.03}$ & 
	$ 1.45 \pm 0.08 \pm 0.05 _{-0.05}^{+0.09}$ & 
	$ 2.01 \pm 0.13 \pm 0.07 _{-0.13}^{+0.10}$ & 
	$ 2.23 \pm 0.15 \pm 0.08 _{-0.22}^{+0.07}$ & 
	$ 3.67 \pm 0.21 \pm 0.12 _{-0.63}^{+0.00}$ & 
	$ 8.42 \pm 1.38 \pm 0.45 _{-1.49}^{+0.00}$  
	 \\[1.01ex]%
12--24 &  
	$ 0.40 \pm 0.04 \pm 0.02 _{-0.00}^{+0.03}$ & 
	$ 1.39 \pm 0.15 \pm 0.08 _{-0.04}^{+0.09}$ & 
	$ 1.77 \pm 0.26 \pm 0.13 _{-0.12}^{+0.09}$ & 
	$ 3.37 \pm 0.30 \pm 0.22 _{-0.32}^{+0.11}$ & 
	$ 3.53 \pm 0.37 \pm 0.21 _{-0.55}^{+0.00}$ & 
	--	 
	 \\[1.01ex]%
\hline 
\end{tabular} 
\caption{Average of relative $\Dzero$, $\Dplus$ and $\Dstar$ meson yields for the sum of particles and antiparticles in several multiplicity and $\pt$ intervals for \pPb~collisions at~$\sqrtsNN=5.02$~TeV as a function of the relative charged-particle multiplicity at central rapidity. The values are reported together with their uncertainties, which are quoted in the order: statistical, systematic and feed-down contribution uncertainties. The yields reported here are per non-single diffractive event. The global normalisation uncertainty of $3.1\%$ is not shown. 
\label{tab:DAverageMult} 
} 
 
\end{center} 
} 
\end{table} 
 
\end{landscape} 

\begin{landscape}

\begin{table}[htp] 
\begin{center} 
\begin{tabular}{lccccc} 
\hline 
&  \multicolumn{5}{c}{$\Nvzero \big/ \langle \Nvzero \rangle$} \\[1.01ex] & 
	 $0.48\pm 0.02$  
	& $1.32 \pm 0.07 $  
	& $1.81 \pm 0.09$   
	&$ 2.36 \pm 0.12 $ 
	& $3.29 \pm 0.16$ \\ 
	&&&& \multicolumn{2}{c}{$(2.72 \pm 0.14)$}  \\ 
\hline 
$\pt$~($\gevc$)&  \multicolumn{5}{c}{${\rm d}^2N/{\rm d}y{\rm d}p_{\rm T} \big/ \langle {\rm d}^2N/{\rm d}y{\rm d}p_{\rm T} \rangle $} \\[1.01ex] 
1--2 &  
	$ 0.55\pm 0.05 \pm 0.02 _{-0.00}^{+0.04}$ & 
	$ 1.57\pm 0.19 \pm 0.06 _{-0.05}^{+0.09}$ & 
	$ 1.92\pm 0.28 \pm 0.08 _{-0.12}^{+0.09}$ & 
\multicolumn{2}{c}{$ (2.69\pm 0.39 \pm 0.14 _{-0.31}^{+0.10})$} 
	 \\[1.01ex]
2--4 &  
	$ 0.52\pm 0.01 \pm 0.01 _{-0.00}^{+0.03}$ & 
	$ 1.47\pm 0.05 \pm 0.04 _{-0.04}^{+0.08}$ & 
	$ 1.91\pm 0.07 \pm 0.05 _{-0.10}^{+0.08}$ & 
	$ 2.50\pm 0.11 \pm 0.07 _{-0.20}^{+0.07}$ & 
	$ 3.25\pm 0.17 \pm 0.09 _{-0.45}^{+0.00}$  
	 \\[1.01ex]%
4--8 &  
	$ 0.51\pm 0.01 \pm 0.01 _{-0.00}^{+0.03}$ & 
	$ 1.59\pm 0.04 \pm 0.03 _{-0.04}^{+0.08}$ & 
	$ 1.93\pm 0.06 \pm 0.04 _{-0.10}^{+0.08}$ & 
	$ 2.43\pm 0.09 \pm 0.05 _{-0.19}^{+0.06}$ & 
	$ 3.02\pm 0.14 \pm 0.07 _{-0.42}^{+0.00}$  
	 \\[1.01ex]%
8--12 &  
	$ 0.55\pm 0.02 \pm 0.02 _{-0.00}^{+0.04}$ & 
	$ 1.60\pm 0.08 \pm 0.05 _{-0.05}^{+0.10}$ & 
	$ 1.84\pm 0.12 \pm 0.06 _{-0.13}^{+0.09}$ & 
	$ 2.62\pm 0.17 \pm 0.08 _{-0.27}^{+0.09}$ & 
	$ 3.03\pm 0.27 \pm 0.10 _{-0.53}^{+0.00}$  
	 \\[1.01ex]%
 
12--24 &  
	$ 0.56\pm 0.04 \pm 0.03 _{-0.00}^{+0.04}$ & 
	$ 1.53\pm 0.15 \pm 0.07 _{-0.05}^{+0.09}$ & 
	$ 2.22\pm 0.22 \pm 0.11 _{-0.14}^{+0.11}$ & 
	 \multicolumn{2}{c}{$ (2.27\pm 0.25 \pm 0.13 _{-0.24}^{+0.08})$} 
	 \\[1.01ex]%

\hline 
\end{tabular} 
 
\end{center} 
\caption{Average of relative $\Dzero$, $\Dplus$ and $\Dstar$ meson yields for the sum of particles and antiparticles in several multiplicity and $\pt$ intervals for \pPb~collisions at~$\sqrtsNN=5.02$~TeV as a function of the relative average multiplicity in the VZERO detector, $\Nvzero \big/ \langle \Nvzero \rangle$. The uncertainties are shown in the following order: statistical, systematic, and feed-down contribution uncertainties. The global normalisation uncertainty of $3.1\%$ is not shown.  
The yields reported here are normalised to the non-single diffractive cross section. For $1 < \pt < 2$~$\gevc$ and $12 < \pt < 24$~$\gevc$, the final two multiplicity intervals are merged, and the results are shown in parentheses. 
\label{tab:DvsNvzero} 
} 
 
\end{table}

\end{landscape} 

\section{The ALICE Collaboration}
\label{app:collab}
\input{Alice_Authorlist_2015-Nov-20.tex}  

\end{document}

%% file: acknowledgements.tex

The ALICE Collaboration would like to thank all its engineers and technicians for their invaluable contributions to the construction of the experiment and the CERN accelerator teams for the outstanding performance of the LHC complex.
The ALICE Collaboration gratefully acknowledges the resources and support provided by all Grid centres and the Worldwide LHC Computing Grid (WLCG) collaboration.
The ALICE Collaboration acknowledges the following funding agencies for their support in building and
running the ALICE detector:
State Committee of Science,  World Federation of Scientists (WFS)
and Swiss Fonds Kidagan, Armenia;
Conselho Nacional de Desenvolvimento Cient\'{\i}fico e Tecnol\'{o}gico (CNPq), Financiadora de Estudos e Projetos (FINEP),
Funda\c{c}\~{a}o de Amparo \`{a} Pesquisa do Estado de S\~{a}o Paulo (FAPESP);
National Natural Science Foundation of China (NSFC), the Chinese Ministry of Education (CMOE)
and the Ministry of Science and Technology of China (MSTC);
Ministry of Education and Youth of the Czech Republic;
Danish Natural Science Research Council, the Carlsberg Foundation and the Danish National Research Foundation;
The European Research Council under the European Community's Seventh Framework Programme;
Helsinki Institute of Physics and the Academy of Finland;
French CNRS-IN2P3, the `Region Pays de Loire', `Region Alsace', `Region Auvergne' and CEA, France;
German Bundesministerium fur Bildung, Wissenschaft, Forschung und Technologie (BMBF) and the Helmholtz Association;
General Secretariat for Research and Technology, Ministry of Development, Greece;
National Research, Development and Innovation Office (NKFIH), Hungary;
Department of Atomic Energy and Department of Science and Technology of the Government of India;
Istituto Nazionale di Fisica Nucleare (INFN) and Centro Fermi -
Museo Storico della Fisica e Centro Studi e Ricerche ``Enrico Fermi'', Italy;
Japan Society for the Promotion of Science (JSPS) KAKENHI and MEXT, Japan;
Joint Institute for Nuclear Research, Dubna;
National Research Foundation of Korea (NRF);
Consejo Nacional de Cienca y Tecnologia (CONACYT), Direccion General de Asuntos del Personal Academico(DGAPA), M\'{e}xico, Amerique Latine Formation academique - 
European Commission~(ALFA-EC) and the EPLANET Program~(European Particle Physics Latin American Network);
Stichting voor Fundamenteel Onderzoek der Materie (FOM) and the Nederlandse Organisatie voor Wetenschappelijk Onderzoek (NWO), Netherlands;
Research Council of Norway (NFR);
National Science Centre, Poland;
Ministry of National Education/Institute for Atomic Physics and National Council of Scientific Research in Higher Education~(CNCSI-UEFISCDI), Romania;
Ministry of Education and Science of Russian Federation, Russian
Academy of Sciences, Russian Federal Agency of Atomic Energy,
Russian Federal Agency for Science and Innovations and The Russian
Foundation for Basic Research;
Ministry of Education of Slovakia;
Department of Science and Technology, South Africa;
Centro de Investigaciones Energeticas, Medioambientales y Tecnologicas (CIEMAT), E-Infrastructure shared between Europe and Latin America (EELA), 
Ministerio de Econom\'{i}a y Competitividad (MINECO) of Spain, Xunta de Galicia (Conseller\'{\i}a de Educaci\'{o}n),
Centro de Aplicaciones Tecnológicas y Desarrollo Nuclear (CEA\-DEN), Cubaenerg\'{\i}a, Cuba, and IAEA (International Atomic Energy Agency);
Swedish Research Council (VR) and Knut $\&$ Alice Wallenberg
Foundation (KAW);
Ukraine Ministry of Education and Science;
United Kingdom Science and Technology Facilities Council (STFC);
The United States Department of Energy, the United States National
Science Foundation, the State of Texas, and the State of Ohio;
Ministry of Science, Education and Sports of Croatia and  Unity through Knowledge Fund, Croatia;
Council of Scientific and Industrial Research (CSIR), New Delhi, India;
Pontificia Universidad Cat\'{o}lica del Per\'{u}.

%% file: Alice_Authorlist_2015-Nov-20.tex
\bigskip 
\begin{flushleft}

J.~Adam$^{\rm 40}$, 
D.~Adamov\'{a}$^{\rm 84}$, 
M.M.~Aggarwal$^{\rm 88}$, 
G.~Aglieri Rinella$^{\rm 36}$, 
M.~Agnello$^{\rm 110}$, 
N.~Agrawal$^{\rm 48}$, 
Z.~Ahammed$^{\rm 132}$, 
S.U.~Ahn$^{\rm 68}$, 
S.~Aiola$^{\rm 136}$, 
A.~Akindinov$^{\rm 58}$
% ...

\bigskip

\bigskip 

\textbf{\Large Affiliation Notes}

\bigskip 

% ...
$^{\rm IV}$ Also at: M.V. Lomonosov Moscow State University, D.V. Skobeltsyn Institute of Nuclear, Physics, Moscow, Russia

\bigskip

\bigskip 

\textbf{\Large Collaboration Institutes}

\bigskip 

% ...
$^{36}$ European Organization for Nuclear Research (CERN), Geneva, Switzerland\\
% ...
$^{40}$ Faculty of Nuclear Sciences and Physical Engineering, Czech Technical University in Prague, Prague, Czech Republic\\
% ...
$^{48}$ Indian Institute of Technology Bombay (IIT), Mumbai, India\\
% ...
$^{58}$ Institute for Theoretical and Experimental Physics, Moscow, Russia\\
% ...
$^{68}$ Korea Institute of Science and Technology Information, Daejeon, South Korea\\
% ...
$^{84}$ Nuclear Physics Institute, Academy of Sciences of the Czech Republic, \v{R}e\v{z} u Prahy, Czech Republic\\
% ...
$^{88}$ Physics Department, Panjab University, Chandigarh, India\\
% ...
$^{110}$ Sezione INFN, Turin, Italy\\
% ...
$^{132}$ Variable Energy Cyclotron Centre, Kolkata, India\\
% ...
$^{136}$ Yale University, New Haven, Connecticut, United States\\
% ...

\bigskip 

\end{flushleft} 